\documentclass[aps,preprint]{revtex4}%
\usepackage{amsfonts}
\usepackage{amsmath}
\usepackage{amssymb}
\usepackage{graphicx}%
\begin{document}
\preprint{ }
\title[Tripartite loss model ]{Tripartite loss model for Mach-Zehnder interferometers with application to
phase sensitivity : Complete expressions for measurement operator mean values,
variances, and cross correlations}
\author{A. D. Parks, S. E. Spence, and J. E. Troupe}
\affiliation{Quantum Processing Group, Electromagnetic and Sensor Systems Department, Naval
Surface Warfare Center, Dahlgren, Virginia 22448 }
\affiliation{Center for Quantum Studies, George Mason University, Fairfax, Virginia 22030}

\begin{abstract}
A generalized analytical tripartite loss model is posited for Mach-Zehnder
interferometer (MZI) phase sensitivity which is valid for both arbitrary
photon input states and arbitrary system environmental states. This model is
shown to subsume the phase sensitivity models for the lossless MZI and the
ground state MZI. It can be employed to develop specialized models useful for
estimating phase sensitivities, as well as for performing associated design
trade-off analyses, for MZIs which operate in environmental regimes that are
not contained within the ground state MZI's envelope of validity. As a simple
illustration of its utility, the model is used to develop phase sensitivity
expressions for an MZI with "excited" internal arms and an MZI with "excited"
output channels. These expressions yield a conditional relationship between
the expected number of photons entering an MZI and its efficiency parameters
which - when satisfied - predicts an enhanced phase sensitivity for the MZI
with "excited" output channels relative to that for the MZI with "excited"
internal arms.

\end{abstract}
\startpage{1}
\endpage{102}
\maketitle

\section{INTRODUCTION}

A loss model for Mach-Zehnder interferometers (MZIs) has recently been
reported by us in the literature\cite{PSTR} (hereafter referred to as PI).
This model is used to develop a phase sensitivity expression for arbitrary
photon input states that is generally applicable for lossy MZIs employing
subunit efficiency homodyne detection schemes. In particular, it is shown
there that this phase sensitivity $\Delta^{2}\varphi$ is given by%
\begin{equation}
\Delta^{2}\varphi=\frac{\left[
\begin{array}
[c]{c}%
\kappa_{\alpha}^{2}\Delta^{2}C_{\alpha}+\kappa_{\gamma}^{2}\Delta^{2}%
C_{\gamma}+\kappa_{\varepsilon}^{2}\Delta^{2}C_{\varepsilon}+\kappa
_{\alpha,\gamma}^{2}\Delta^{2}C_{\alpha,\gamma}+\kappa_{\alpha,\varepsilon
}^{2}\Delta^{2}C_{\alpha,\varepsilon}\\
+\kappa_{\gamma,\varepsilon}^{2}\Delta^{2}C_{\gamma,\varepsilon}%
+\kappa_{\alpha}\kappa_{\alpha,\gamma}\left\langle \widehat{D}_{\alpha
;\alpha,\gamma}\right\rangle +\kappa_{\alpha}\kappa_{\alpha,\varepsilon
}\left\langle \widehat{D}_{\alpha;\alpha,\varepsilon}\right\rangle
+\kappa_{\gamma}\kappa_{\alpha,\gamma}\left\langle \widehat{D}_{\gamma
;\alpha,\gamma}\right\rangle \\
+\kappa_{\gamma}\kappa_{\gamma,\varepsilon}\left\langle \widehat{D}%
_{\gamma;\gamma,\varepsilon}\right\rangle +\kappa_{\varepsilon}\kappa
_{\alpha,\varepsilon}\left\langle \widehat{D}_{\varepsilon;\alpha,\varepsilon
}\right\rangle +\kappa_{\varepsilon}\kappa_{\gamma,\varepsilon}\left\langle
\widehat{D}_{\varepsilon;\gamma,\varepsilon}\right\rangle +\kappa
_{\alpha,\gamma}\kappa_{\alpha,\varepsilon}\left\langle \widehat{D}%
_{\alpha,\gamma;\alpha,\varepsilon}\right\rangle \\
+\kappa_{\alpha,\gamma}\kappa_{\gamma,\varepsilon}\left\langle \widehat
{D}_{\alpha,\gamma;\gamma,\varepsilon}\right\rangle +\kappa_{\alpha
,\varepsilon}\kappa_{\gamma,\varepsilon}\left\langle \widehat{D}%
_{\alpha,\varepsilon;\gamma,\varepsilon}\right\rangle
\end{array}
\right]  }{\left\vert \kappa_{\alpha}\frac{\partial\left\langle \widehat
{C}_{\alpha}\right\rangle }{\partial\varphi}+\kappa_{\alpha,\gamma}%
\frac{\partial\left\langle \widehat{C}_{\alpha,\gamma}\right\rangle }%
{\partial\varphi}+\kappa_{\alpha,\varepsilon}\frac{\partial\left\langle
\widehat{C}_{\alpha,\varepsilon}\right\rangle }{\partial\varphi}\right\vert
^{2}}, \label{D2PHI}%
\end{equation}
where $\varphi$ is the associated phase angle to be measured; the subscripts
$x\in\{\alpha,\gamma,\varepsilon\}$ refer to the three regions used in the
model ($0<x\leq1$ is the value of the efficiency parameter for region $x$);
and the subscripted $\kappa$'s are constants which depend upon the regional
efficiency parameters. The quantities $\Delta^{2}C_{x}=\left\langle
\widehat{C}_{x}^{2}\right\rangle -\left\langle \widehat{C}_{x}\right\rangle
^{2}$ and $\left\langle \widehat{D}_{x;y}\right\rangle =\left\langle
\widehat{C}_{x}\widehat{C}_{y}+\widehat{C}_{y}\widehat{C}_{x}\right\rangle
-2\left\langle \widehat{C}_{x}\right\rangle \left\langle \widehat{C}%
_{y}\right\rangle $ are measurement operator variances and cross correlations
(the operator $\widehat{C}$ measures the difference in the number of photons
exiting the output ports), respectively, and $\left\langle \widehat
{Y}\right\rangle =\left\langle \Psi\right\vert $ $\widehat{Y}\left\vert
\Psi\right\rangle $ is the mean value of the operator $\widehat{Y}$ for the
system state $\left\vert \Psi\right\rangle =\left\vert \psi_{a_{in},b_{in}%
}\right\rangle \left\vert \psi_{c}\right\rangle \left\vert \psi_{d}%
\right\rangle \left\vert \psi_{e}\right\rangle \left\vert \psi_{f}%
\right\rangle \left\vert \psi_{g}\right\rangle \left\vert \psi_{h}%
\right\rangle $. Here $\left\vert \psi_{a_{in},b_{in}}\right\rangle $ is the
preloss input state and $\left\vert \psi_{x}\right\rangle $, $x\in
\{c,d,e,f,g,h\}$, is the normalized system environmental state associated with
the system environmental annihilation operator $\widehat{x}$. Precise
definitions for the subscripted $\kappa$'s and $\widehat{C}_{x}$ operators
appearing above are provided by eqs.(1) - (5) in PI. The reader is referred to
section II in PI for a description of the\ physical assumptions and the
regional architecture upon which the model is based.

The primary focus of PI was the development of the phase sensitivity
expression for the ground state MZI. Such an MZI is defined in terms of our
model as one which exhibits loss via non-unit regional efficiency parameter
values and for which all system environmental states are vacuum states, i.e.
$\left\vert \psi_{x}\right\rangle =\left\vert 0\right\rangle $, $x\in
\{c,d,e,f,g,h\}$, so that $\left\vert \Psi\right\rangle =\left\vert
\psi_{a_{in},b_{in}}\right\rangle \left\vert 0\right\rangle \left\vert
0\right\rangle \left\vert 0\right\rangle \left\vert 0\right\rangle \left\vert
0\right\rangle \left\vert 0\right\rangle $. In this case, the ground state
phase sensitivity $\Delta^{2}\varphi_{gs}$ is given by eq.(\ref{D2PHI}) with
all but the first, fourth, and fifth terms in the numerator and the first term
in the denominator set equal to zero (see eqs.(11) and (12) in PI).

This expression for $\Delta^{2}\varphi_{gs}$ is a useful result because - as
discussed in PI - it represents to a good approximation MZI phase sensitivity
for a wide range of MZI environmental temperatures for frequencies in the
near-IR to the near-UV region of the electromagnetic spectrum. However, for
certain MZI applications\ it may be important to model the phase sensitivity
for conditions where the system environmental states are not vacuum states,
i.e. the MZI is not in its ground state, so that eq.(12) in PI does not apply.
When using eq.(\ref{D2PHI}) to model $\Delta^{2}\varphi$ for such cases, it is
necessary to have the complete analytical expressions for each term appearing
in the right hand side of this equation.

The purpose of this article is to extend the results of PI by providing the
expressions for these terms so that eq.(\ref{D2PHI}) is a complete model that
may be more generally useful for MZI phase sensitivity analyses involving
system environmental regimes which are not contained within the envelope of
validity for the ground state model. Complete expressions for the partial
phase derivatives of the measurement operator mean values, the measurement
operator variances, and the measurement operator cross correlations appearing
in the right hand side of eq.(\ref{D2PHI}) are provided, respectively, in the
following three sections of this paper. These expressions are validated in the
final section of this paper by demonstrating that - when used in conjunction
with eq.(\ref{D2PHI}) - they yield the phase sensitivities for the lossless
MZI and the ground state MZI previously developed in PI. As an additional
illustration of its utility as an analytical tool, the model is employed to
provide the phase sensitivity expressions for an MZI with "excited" internal
arms, i.e. an "excited" $\gamma$-region configuration, and\ for an MZI with
"excited" output channels, i.e. an "excited" $\varepsilon$-region
configuration. These expressions are used to establish an associated
conditional relationship between the expected number of photons entering an
MZI and its regional efficiency parameter values. When this condition\ is
satisfied, then the phase sensitivity of the "excited" $\gamma$-region
configuration is more degraded than that of the "excited" $\varepsilon$-region configuration.

\section{PARTIAL PHASE\ DERIVATIVES\ OF\ MEASUREMENT\ OPERATOR\ MEAN\ VALUES}

The expressions for the measurement operator mean values needed to evaluate
the partial derivatives appearing in the denominator of eq.(\ref{D2PHI}) are%
\begin{equation}
\left\langle \hat{C}_{\alpha}\right\rangle =4\left\{  \left(  \left\langle
\hat{\rho}_{b}^{\dagger}\hat{\rho}_{b}\right\rangle -\left\langle \hat{\rho
}_{a}^{\dagger}\hat{\rho}_{a}\right\rangle \right)  \cos\varphi-\left(
\left\langle \hat{\rho}_{a}^{\dagger}\hat{\rho}_{b}\right\rangle +\left\langle
\hat{\rho}_{b}^{\dagger}\hat{\rho}_{a}\right\rangle \right)  \sin
\varphi\right\}  , \label{CA}%
\end{equation}%
\begin{equation}
\left\langle \hat{C}_{\alpha,\gamma}\right\rangle =2\left\{
\begin{array}
[c]{c}%
\left(  \left\langle \hat{\rho}_{b}^{\dagger}\hat{f}\right\rangle
+\left\langle \hat{f}^{\dagger}\hat{\rho}_{b}\right\rangle \right)  +i\left(
\left\langle \hat{\rho}_{a}^{\dagger}\hat{f}\right\rangle -\left\langle
\hat{f}^{\dagger}\hat{\rho}_{a}\right\rangle \right) \\
-i\left(  \left\langle \hat{\rho}_{b}^{\dagger}\hat{e}\right\rangle
e^{-i\varphi}-\left\langle \hat{e}^{\dagger}\hat{\rho}_{b}\right\rangle
e^{i\varphi}\right)  -\left(  \left\langle \hat{\rho}_{a}^{\dagger}\hat
{e}\right\rangle e^{-i\varphi}+\left\langle \hat{e}^{\dagger}\hat{\rho}%
_{a}\right\rangle e^{i\varphi}\right)
\end{array}
\right\}  , \label{CAG}%
\end{equation}
and%
\begin{align}
\left\langle \hat{C}_{\alpha,\varepsilon}\right\rangle  &  =\left\{
\left\langle \hat{\rho}_{a}^{\dagger}\hat{g}\right\rangle \left(
1-e^{-i\varphi}\right)  +\left\langle \hat{g}^{\dagger}\hat{\rho}%
_{a}\right\rangle \left(  1-e^{i\varphi}\right)  \right\}  +\left\{
\left\langle \hat{\rho}_{b}^{\dagger}\hat{h}\right\rangle \left(
1-e^{-i\varphi}\right)  +\left\langle \hat{h}^{\dagger}\hat{\rho}%
_{b}\right\rangle \left(  1-e^{i\varphi}\right)  \right\} \label{CAE}\\
&  +i\left\{  \left\langle \hat{\rho}_{a}^{\dagger}\hat{h}\right\rangle
\left(  1+e^{-i\varphi}\right)  -\left\langle \hat{h}^{\dagger}\hat{\rho}%
_{a}\right\rangle \left(  1+e^{i\varphi}\right)  \right\}  -i\left\{
\left\langle \hat{\rho}_{b}^{\dagger}\hat{g}\right\rangle \left(
1+e^{-i\varphi}\right)  -\left\langle \hat{g}^{\dagger}\hat{\rho}%
_{b}\right\rangle \left(  1+e^{i\varphi}\right)  \right\}  ,\nonumber
\end{align}
where%
\[
\widehat{\rho}_{a}=\sqrt{\alpha}\widehat{a}_{in}+\sqrt{1-\alpha}\widehat{c}%
\]
and%
\[
\widehat{\rho}_{b}=\sqrt{\alpha}\widehat{b}_{in}+\sqrt{1-\alpha}\widehat{d}.
\]
Here $0<\alpha\leq1$ is the efficiency parameter for the $\alpha$ region of
the model, $\widehat{a}_{in}$ and $\widehat{b}_{in}$ are the input port
annihilation operators, $\widehat{c}$ and $\widehat{d}$ are the $\alpha$
region\ environmental annihilation operators, $\widehat{e}$ and $\widehat{f}$
are the $\gamma$ region environmental annihilation operators, and $\widehat
{g}$ and $\widehat{h}$ are the $\varepsilon$ region environmental annihilation operators.

The partial derivatives appearing in eq.(\ref{D2PHI}) are readily obtained as
follows from eqs. (\ref{CA}) - (\ref{CAE}) :%
\begin{equation}
\frac{\partial\left\langle \widehat{C}_{\alpha}\right\rangle }{\partial
\varphi}=-4\{\left(  \left\langle \widehat{\rho}_{b}^{\dag}\widehat{\rho}%
_{b}\right\rangle -\left\langle \widehat{\rho}_{a}^{\dag}\widehat{\rho}%
_{a}\right\rangle \right)  \sin\varphi+\left(  \left\langle \widehat{\rho}%
_{a}^{\dag}\widehat{\rho}_{b}\right\rangle +\left\langle \widehat{\rho}%
_{b}^{\dag}\widehat{\rho}_{a}\right\rangle \right)  \cos\varphi\}, \label{DCA}%
\end{equation}%
\[
\frac{\partial\left\langle \widehat{C}_{\alpha,\gamma}\right\rangle }%
{\partial\varphi}=-2\left\{  \left(  \left\langle \widehat{\rho}_{b}^{\dag
}\widehat{e}\right\rangle -i\left\langle \widehat{\rho}_{a}^{\dag}\widehat
{e}\right\rangle \right)  e^{-i\varphi}+\left(  \left\langle \widehat{e}%
^{\dag}\widehat{\rho}_{b}\right\rangle +i\left\langle \widehat{e}^{\dag
}\widehat{\rho}_{a}\right\rangle \right)  e^{i\varphi}\right\}  ,
\]
and%
\[
\frac{\partial\left\langle \widehat{C}_{\alpha,\varepsilon}\right\rangle
}{\partial\varphi}=\left\{
\begin{array}
[c]{c}%
\left[  \left(  \left\langle \widehat{\rho}_{a}^{\dag}\widehat{h}\right\rangle
-\left\langle \widehat{\rho}_{b}^{\dag}\widehat{g}\right\rangle \right)
+i\left(  \left\langle \widehat{\rho}_{a}^{\dag}\widehat{g}\right\rangle
+\left\langle \widehat{\rho}_{b}^{\dag}\widehat{h}\right\rangle \right)
\right]  e^{-i\varphi}\\
+\left[  \left(  \left\langle \widehat{h}^{\dag}\widehat{\rho}_{a}%
\right\rangle -\left\langle \widehat{g}^{\dag}\widehat{\rho}_{b}\right\rangle
\right)  -i\left(  \left\langle \widehat{g}^{\dag}\widehat{\rho}%
_{a}\right\rangle +\left\langle \widehat{h}^{\dag}\widehat{\rho}%
_{b}\right\rangle \right)  \right]  e^{i\varphi}%
\end{array}
\right\}  .
\]

\section{MEASUREMENT\ OPERATOR\ VARIANCES}

Expressions for the measurement operator variances needed for the evaluation
of the numerator of eq.(\ref{D2PHI}) are
\begin{equation}
\Delta^{2}C_{\alpha}=16\left[
\begin{array}
[c]{c}%
\left(  \left\langle \left(  \hat{\rho}_{b}^{\dagger}\hat{\rho}_{b}-\hat{\rho
}_{a}^{\dagger}\hat{\rho}_{a}\right)  ^{2}\right\rangle -\left\langle
\hat{\rho}_{b}^{\dagger}\hat{\rho}_{b}-\hat{\rho}_{a}^{\dagger}\hat{\rho}%
_{a}\right\rangle ^{2}\right)  \cos^{2}\varphi+\\
\left(  \left\langle \left(  \hat{\rho}_{a}^{\dagger}\hat{\rho}_{b}+\hat{\rho
}_{b}^{\dagger}\hat{\rho}_{a}\right)  ^{2}\right\rangle -\left\langle
\hat{\rho}_{a}^{\dagger}\hat{\rho}_{b}+\hat{\rho}_{b}^{\dagger}\hat{\rho}%
_{a}\right\rangle ^{2}\right)  \sin^{2}\varphi-\\
\left(
\begin{array}
[c]{c}%
\left\langle \left\{  \left(  \hat{\rho}_{b}^{\dagger}\hat{\rho}_{b}-\hat
{\rho}_{a}^{\dagger}\hat{\rho}_{a}\right)  ,\left(  \hat{\rho}_{a}^{\dagger
}\hat{\rho}_{b}+\hat{\rho}_{b}^{\dagger}\hat{\rho}_{a}\right)  \right\}
\right\rangle \\
-2\left\langle \hat{\rho}_{b}^{\dagger}\hat{\rho}_{b}-\hat{\rho}_{a}^{\dagger
}\hat{\rho}_{a}\right\rangle \left\langle \hat{\rho}_{a}^{\dagger}\hat{\rho
}_{b}+\hat{\rho}_{b}^{\dagger}\hat{\rho}_{a}\right\rangle
\end{array}
\right)  \sin\varphi\cos\varphi
\end{array}
\right]  , \label{VARIN}%
\end{equation}
where the term enclosed in curley braces is the anti-commutator defined by
$\left\{  \widehat{X},\widehat{Y}\right\}  =\widehat{X}\widehat{Y}+\widehat
{Y}\widehat{X}$;%
\begin{equation}
\Delta^{2}C_{\gamma}=4\left[
\begin{array}
[c]{c}%
\left\langle \hat{e}^{\dagger}\hat{e}\hat{f}\hat{f}^{\dagger}\right\rangle
+\left\langle \hat{e}\hat{e}^{\dagger}\hat{f}^{\dagger}\hat{f}\right\rangle
+\left\langle \hat{e}^{\dagger}\hat{f}\right\rangle ^{2}+\left\langle \hat
{e}\hat{f}^{\dagger}\right\rangle ^{2}\\
-\left\langle \hat{e}^{\dagger}\hat{e}^{\dagger}\hat{f}\hat{f}\right\rangle
-\left\langle \hat{e}\hat{e}\hat{f}^{\dagger}\hat{f}^{\dagger}\right\rangle
-2\left\langle \hat{e}^{\dagger}\hat{f}\right\rangle \left\langle \hat{e}%
\hat{f}^{\dagger}\right\rangle
\end{array}
\right]  ; \label{VARFB}%
\end{equation}%
\[
\Delta^{2}C_{\varepsilon}=\left[
\begin{array}
[c]{c}%
\left(  \left\langle \hat{g}^{\dagger}\hat{g}\hat{g}^{\dagger}\hat
{g}\right\rangle -\left\langle \hat{g}^{\dagger}\hat{g}\right\rangle
^{2}\right)  +\left(  \left\langle \hat{h}^{\dagger}\hat{h}\hat{h}^{\dagger
}\hat{h}\right\rangle -\left\langle \hat{h}^{\dagger}\hat{h}\right\rangle
^{2}\right) \\
-2\left(  \left\langle \hat{g}^{\dagger}\hat{g}\hat{h}^{\dagger}\hat
{h}\right\rangle -\left\langle \hat{g}^{\dagger}\hat{g}\right\rangle
\left\langle \hat{h}^{\dagger}\hat{h}\right\rangle \right)
\end{array}
\right]  ;
\]%
\begin{equation}
\Delta^{2}C_{\alpha,\gamma}=4\left[
\begin{array}
[c]{c}%
\left\langle \hat{F}_{1}\hat{e}\hat{e}\right\rangle +\left\langle \hat{F}%
_{1}^{\dagger}\hat{e}^{\dagger}\hat{e}^{\dagger}\right\rangle +\left\langle
\hat{F}_{2}\hat{e}^{\dagger}\hat{e}\right\rangle +\left\langle \hat{F}_{3}%
\hat{e}\hat{e}^{\dagger}\right\rangle -\left\langle \hat{F}_{4}\hat{f}\hat
{f}\right\rangle -\left\langle \hat{F}_{4}^{\dagger}\hat{f}^{\dagger}\hat
{f}^{\dagger}\right\rangle +\\
\left\langle \hat{F}_{5}\hat{f}^{\dagger}\hat{f}\right\rangle +\left\langle
\hat{F}_{6}\hat{f}\hat{f}^{\dagger}\right\rangle +\left\langle \hat{F}_{7}%
\hat{e}\hat{f}\right\rangle +\left\langle \hat{F}_{7}^{\dagger}\hat
{e}^{\dagger}\hat{f}^{\dagger}\right\rangle -\left\langle \hat{F}_{8}\hat
{e}^{\dagger}\hat{f}\right\rangle -\left\langle \hat{F}_{8}^{\dagger}\hat
{e}\hat{f}^{\dagger}\right\rangle -\\
\left\langle \hat{\rho}_{b}^{\dagger}\hat{f}\right\rangle ^{2}-\left\langle
\hat{\rho}_{b}\hat{f}^{\dagger}\right\rangle ^{2}+\left\langle \hat{\rho}%
_{a}^{\dagger}\hat{f}\right\rangle ^{2}+\left\langle \hat{\rho}_{a}\hat
{f}^{\dagger}\right\rangle ^{2}-\\
2\left(  \left\langle \hat{\rho}_{b}^{\dagger}\hat{f}\right\rangle
\left\langle \hat{\rho}_{b}\hat{f}^{\dagger}\right\rangle +\left\langle
\hat{\rho}_{a}^{\dagger}\hat{f}\right\rangle \left\langle \hat{\rho}_{a}%
\hat{f}^{\dagger}\right\rangle +\left\langle \hat{\rho}_{b}^{\dagger}\hat
{e}\right\rangle \left\langle \hat{\rho}_{b}\hat{e}^{\dagger}\right\rangle
+\left\langle \hat{\rho}_{a}^{\dagger}\hat{e}\right\rangle \left\langle
\hat{\rho}_{a}\hat{e}^{\dagger}\right\rangle \right)  -\\
\left(  \left\langle \hat{\rho}_{a}^{\dagger}\hat{e}\right\rangle
^{2}-\left\langle \hat{\rho}_{b}^{\dagger}\hat{e}\right\rangle ^{2}\right)
e^{-2i\varphi}-\left(  \left\langle \hat{\rho}_{a}\hat{e}^{\dagger
}\right\rangle ^{2}-\left\langle \hat{\rho}_{b}\hat{e}^{\dagger}\right\rangle
^{2}\right)  e^{2i\varphi}-\\
2\left(  \left\langle \hat{\rho}_{a}^{\dagger}\hat{f}\right\rangle
\left\langle \hat{\rho}_{b}^{\dagger}\hat{e}\right\rangle -\left\langle
\hat{\rho}_{a}\hat{f}^{\dagger}\right\rangle \left\langle \hat{\rho}%
_{b}^{\dagger}\hat{e}\right\rangle -\left\langle \hat{\rho}_{b}^{\dagger}%
\hat{f}\right\rangle \left\langle \hat{\rho}_{a}^{\dagger}\hat{e}\right\rangle
-\left\langle \hat{\rho}_{b}\hat{f}^{\dagger}\right\rangle \left\langle
\hat{\rho}_{a}^{\dagger}\hat{e}\right\rangle \right)  e^{-i\varphi}-\\
2\left(  \left\langle \hat{\rho}_{a}\hat{f}^{\dagger}\right\rangle
\left\langle \hat{\rho}_{b}\hat{e}^{\dagger}\right\rangle -\left\langle
\hat{\rho}_{a}^{\dagger}\hat{f}\right\rangle \left\langle \hat{\rho}_{b}%
\hat{e}^{\dagger}\right\rangle -\left\langle \hat{\rho}_{b}\hat{f}^{\dagger
}\right\rangle \left\langle \hat{\rho}_{a}\hat{e}^{\dagger}\right\rangle
-\left\langle \hat{\rho}_{b}^{\dagger}\hat{f}\right\rangle \left\langle
\hat{\rho}_{a}\hat{e}^{\dagger}\right\rangle \right)  e^{i\varphi}-\\
2i[\left\langle \hat{\rho}_{a}^{\dagger}\hat{f}\right\rangle \left\langle
\hat{\rho}_{b}^{\dagger}\hat{f}\right\rangle +\left\langle \hat{\rho}%
_{a}^{\dagger}\hat{f}\right\rangle \left\langle \hat{\rho}_{b}\hat{f}%
^{\dagger}\right\rangle +\left\langle \hat{\rho}_{a}\hat{e}^{\dagger
}\right\rangle \left\langle \hat{\rho}_{b}^{\dagger}\hat{e}\right\rangle
-\left\langle \hat{\rho}_{a}\hat{f}^{\dagger}\right\rangle \left\langle
\hat{\rho}_{b}\hat{f}^{\dagger}\right\rangle -\\
\left\langle \hat{\rho}_{a}\hat{f}^{\dagger}\right\rangle \left\langle
\hat{\rho}_{b}^{\dagger}\hat{f}\right\rangle -\left\langle \hat{\rho}%
_{a}^{\dagger}\hat{e}\right\rangle \left\langle \hat{\rho}_{b}\hat{e}%
^{\dagger}\right\rangle +\left\langle \hat{\rho}_{a}^{\dagger}\hat
{e}\right\rangle \left\langle \hat{\rho}_{b}^{\dagger}\hat{e}\right\rangle
e^{-2i\varphi}-\left\langle \hat{\rho}_{a}\hat{e}^{\dagger}\right\rangle
\left\langle \hat{\rho}_{b}\hat{e}^{\dagger}\right\rangle e^{2i\varphi}+\\
\left(  \left\langle \hat{\rho}_{a}\hat{f}^{\dagger}\right\rangle \left\langle
\hat{\rho}_{a}^{\dagger}\hat{e}\right\rangle -\left\langle \hat{\rho}%
_{a}^{\dagger}\hat{f}\right\rangle \left\langle \hat{\rho}_{a}^{\dagger}%
\hat{e}\right\rangle -\left\langle \hat{\rho}_{b}^{\dagger}\hat{f}%
\right\rangle \left\langle \hat{\rho}_{b}^{\dagger}\hat{e}\right\rangle
-\left\langle \hat{\rho}_{b}\hat{f}^{\dagger}\right\rangle \left\langle
\hat{\rho}_{b}^{\dagger}\hat{e}\right\rangle \right)  e^{-i\varphi}+\\
\left(  \left\langle \hat{\rho}_{b}\hat{f}^{\dagger}\right\rangle \left\langle
\hat{\rho}_{b}\hat{e}^{\dagger}\right\rangle +\left\langle \hat{\rho}%
_{b}^{\dagger}\hat{f}\right\rangle \left\langle \hat{\rho}_{b}\hat{e}%
^{\dagger}\right\rangle +\left\langle \hat{\rho}_{a}\hat{f}^{\dagger
}\right\rangle \left\langle \hat{\rho}_{a}\hat{e}^{\dagger}\right\rangle
-\left\langle \hat{\rho}_{a}^{\dagger}\hat{f}\right\rangle \left\langle
\hat{\rho}_{a}\hat{e}^{\dagger}\right\rangle \right)  e^{i\varphi}]
\end{array}
\right]  , \label{VARIF}%
\end{equation}
where
\[
\hat{F}_{1}=\left(  \hat{\rho}_{a}^{\dagger}\hat{\rho}_{a}^{\dagger}-\hat
{\rho}_{b}^{\dagger}\hat{\rho}_{b}^{\dagger}+2i\hat{\rho}_{a}^{\dagger}%
\hat{\rho}_{b}^{\dagger}\right)  e^{-2i\varphi},
\]%
\[
\hat{F}_{2}=\hat{\rho}_{a}\hat{\rho}_{a}^{\dagger}+\hat{\rho}_{b}\hat{\rho
}_{b}^{\dagger}+i\left(  \hat{\rho}_{a}\hat{\rho}_{b}^{\dagger}-\hat{\rho}%
_{a}^{\dagger}\hat{\rho}_{b}\right)  ,
\]%
\[
\hat{F}_{3}=\hat{\rho}_{a}^{\dagger}\hat{\rho}_{a}+\hat{\rho}_{b}^{\dagger
}\hat{\rho}_{b}+i\left(  \hat{\rho}_{a}\hat{\rho}_{b}^{\dagger}-\hat{\rho}%
_{a}^{\dagger}\hat{\rho}_{b}\right)  ,
\]%
\[
\hat{F}_{4}=\hat{\rho}_{a}^{\dagger}\hat{\rho}_{a}^{\dagger}-\hat{\rho}%
_{b}^{\dagger}\hat{\rho}_{b}^{\dagger}-2i\hat{\rho}_{a}^{\dagger}\hat{\rho
}_{b}^{\dagger},
\]%
\[
\hat{F}_{5}=\hat{\rho}_{a}\hat{\rho}_{a}^{\dagger}+\hat{\rho}_{b}\hat{\rho
}_{b}^{\dagger}+i\left(  \hat{\rho}_{a}^{\dagger}\hat{\rho}_{b}-\hat{\rho}%
_{b}^{\dagger}\hat{\rho}_{a}\right)  ,
\]%
\[
\hat{F}_{6}=\hat{\rho}_{a}^{\dagger}\hat{\rho}_{a}+\hat{\rho}_{b}^{\dagger
}\hat{\rho}_{b}+i\left(  \hat{\rho}_{a}^{\dagger}\hat{\rho}_{b}-\hat{\rho}%
_{b}^{\dagger}\hat{\rho}_{a}\right)  ,
\]%
\[
\hat{F}_{7}=-2i\left(  \hat{\rho}_{a}^{\dagger}\hat{\rho}_{a}^{\dagger}%
+\hat{\rho}_{b}^{\dagger}\hat{\rho}_{b}^{\dagger}\right)  e^{-i\varphi},
\]%
\[
\hat{F}_{8}=\left[  2\left(  \hat{\rho}_{a}\hat{\rho}_{b}^{\dagger}+\hat{\rho
}_{a}^{\dagger}\hat{\rho}_{b}\right)  +i\left(  \hat{\rho}_{a}^{\dagger}%
\hat{\rho}_{a}+\hat{\rho}_{a}\hat{\rho}_{a}^{\dagger}-\hat{\rho}_{b}^{\dagger
}\hat{\rho}_{b}-\hat{\rho}_{b}\hat{\rho}_{b}^{\dagger}\right)  \right]
e^{i\varphi};
\]%
\begin{equation}
\Delta^{2}C_{\alpha,\varepsilon}=\left[
\begin{array}
[c]{c}%
\left\langle \hat{R}_{1}\hat{g}\hat{g}\right\rangle +\left\langle \hat{R}%
_{1}^{\dagger}\hat{g}^{\dagger}\hat{g}^{\dagger}\right\rangle +\left\langle
\hat{R}_{2}\hat{g}^{\dagger}\hat{g}\right\rangle +\left\langle \hat{R}_{3}%
\hat{g}\hat{g}^{\dagger}\right\rangle +\left\langle \hat{R}_{4}\hat{h}\hat
{h}\right\rangle +\left\langle \hat{R}_{4}^{\dagger}\hat{h}^{\dagger}\hat
{h}^{\dagger}\right\rangle +\\
\left\langle \hat{R}_{5}\hat{h}^{\dagger}\hat{h}\right\rangle +\left\langle
\hat{R}_{6}\hat{h}\hat{h}^{\dagger}\right\rangle +\left\langle \hat{R}_{7}%
\hat{g}\hat{h}\right\rangle +\left\langle \hat{R}_{7}^{\dagger}\hat
{g}^{\dagger}\hat{h}^{\dagger}\right\rangle +\left\langle \hat{R}_{8}\hat
{g}^{\dagger}\hat{h}\right\rangle +\left\langle \hat{R}_{8}^{\dagger}\hat
{g}\hat{h}^{\dagger}\right\rangle -\\
\left(  \left\langle \hat{\rho}_{a}^{\dagger}\hat{g}\right\rangle
^{2}+\left\langle \hat{\rho}_{b}^{\dagger}\hat{h}\right\rangle ^{2}%
+2\left\langle \hat{\rho}_{a}^{\dagger}\hat{g}\right\rangle \left\langle
\hat{\rho}_{b}^{\dagger}\hat{h}\right\rangle \right)  \left(  1-e^{-i\varphi
}\right)  ^{2}-\\
\left(  \left\langle \hat{\rho}_{a}\hat{g}^{\dagger}\right\rangle
^{2}+\left\langle \hat{\rho}_{b}\hat{h}^{\dagger}\right\rangle ^{2}%
+2\left\langle \hat{\rho}_{a}\hat{g}^{\dagger}\right\rangle \left\langle
\hat{\rho}_{b}\hat{h}^{\dagger}\right\rangle \right)  \left(  1-e^{i\varphi
}\right)  ^{2}+\\
\left(  \left\langle \hat{\rho}_{a}^{\dagger}\hat{h}\right\rangle
^{2}+\left\langle \hat{\rho}_{b}^{\dagger}\hat{g}\right\rangle ^{2}%
-2\left\langle \hat{\rho}_{a}^{\dagger}\hat{h}\right\rangle \left\langle
\hat{\rho}_{b}^{\dagger}\hat{g}\right\rangle \right)  \left(  1+e^{-i\varphi
}\right)  ^{2}+\\
\left(  \left\langle \hat{\rho}_{a}\hat{h}^{\dagger}\right\rangle
^{2}+\left\langle \hat{\rho}_{b}\hat{g}^{\dagger}\right\rangle ^{2}%
-2\left\langle \hat{\rho}_{a}\hat{h}^{\dagger}\right\rangle \left\langle
\hat{\rho}_{b}\hat{g}^{\dagger}\right\rangle \right)  \left(  1+e^{i\varphi
}\right)  ^{2}-\\
2\left(
\begin{array}
[c]{c}%
\left\langle \hat{\rho}_{a}^{\dagger}\hat{g}\right\rangle \left\langle
\hat{\rho}_{a}\hat{g}^{\dagger}\right\rangle +\left\langle \hat{\rho}%
_{b}^{\dagger}\hat{h}\right\rangle \left\langle \hat{\rho}_{b}\hat{h}%
^{\dagger}\right\rangle \\
+\left\langle \hat{\rho}_{a}^{\dagger}\hat{g}\right\rangle \left\langle
\hat{\rho}_{b}\hat{h}^{\dagger}\right\rangle +\left\langle \hat{\rho}_{a}%
\hat{g}^{\dagger}\right\rangle \left\langle \hat{\rho}_{b}^{\dagger}\hat
{h}\right\rangle
\end{array}
\right)  \left(  1-e^{i\varphi}\right)  \left(  1-e^{-i\varphi}\right)  -\\
2\left(
\begin{array}
[c]{c}%
\left\langle \hat{\rho}_{a}^{\dagger}\hat{h}\right\rangle \left\langle
\hat{\rho}_{a}\hat{h}^{\dagger}\right\rangle +\left\langle \hat{\rho}%
_{b}^{\dagger}\hat{g}\right\rangle \left\langle \hat{\rho}_{b}\hat{g}%
^{\dagger}\right\rangle \\
-\left\langle \hat{\rho}_{a}^{\dagger}\hat{h}\right\rangle \left\langle
\hat{\rho}_{b}\hat{g}^{\dagger}\right\rangle -\left\langle \hat{\rho}_{a}%
\hat{h}^{\dagger}\right\rangle \left\langle \hat{\rho}_{b}^{\dagger}\hat
{g}\right\rangle
\end{array}
\right)  \left(  1+e^{i\varphi}\right)  \left(  1+e^{-i\varphi}\right)  -\\
2i[\left(
\begin{array}
[c]{c}%
\left\langle \hat{\rho}_{a}^{\dagger}\hat{g}\right\rangle \left\langle
\hat{\rho}_{a}^{\dagger}\hat{h}\right\rangle +\left\langle \hat{\rho}%
_{b}^{\dagger}\hat{h}\right\rangle \left\langle \hat{\rho}_{a}^{\dagger}%
\hat{h}\right\rangle \\
-\left\langle \hat{\rho}_{a}^{\dagger}\hat{g}\right\rangle \left\langle
\hat{\rho}_{b}^{\dagger}\hat{g}\right\rangle -\left\langle \hat{\rho}%
_{b}^{\dagger}\hat{h}\right\rangle \left\langle \hat{\rho}_{b}^{\dagger}%
\hat{g}\right\rangle
\end{array}
\right)  \left(  1-e^{-i\varphi}\right)  \left(  1+e^{-i\varphi}\right)  +\\
\left(
\begin{array}
[c]{c}%
\left\langle \hat{\rho}_{a}\hat{g}^{\dagger}\right\rangle \left\langle
\hat{\rho}_{b}\hat{g}^{\dagger}\right\rangle +\left\langle \hat{\rho}_{b}%
\hat{h}^{\dagger}\right\rangle \left\langle \hat{\rho}_{b}\hat{g}^{\dagger
}\right\rangle \\
-\left\langle \hat{\rho}_{a}\hat{g}^{\dagger}\right\rangle \left\langle
\hat{\rho}_{a}\hat{h}^{\dagger}\right\rangle -\left\langle \hat{\rho}_{b}%
\hat{h}^{\dagger}\right\rangle \left\langle \hat{\rho}_{a}\hat{h}^{\dagger
}\right\rangle
\end{array}
\right)  \left(  1-e^{i\varphi}\right)  \left(  1+e^{i\varphi}\right)  +\\
\left(
\begin{array}
[c]{c}%
\left\langle \hat{\rho}_{a}^{\dagger}\hat{g}\right\rangle \left\langle
\hat{\rho}_{b}\hat{g}^{\dagger}\right\rangle +\left\langle \hat{\rho}%
_{b}^{\dagger}\hat{h}\right\rangle \left\langle \hat{\rho}_{b}\hat{g}%
^{\dagger}\right\rangle \\
-\left\langle \hat{\rho}_{a}^{\dagger}\hat{g}\right\rangle \left\langle
\hat{\rho}_{a}\hat{h}^{\dagger}\right\rangle -\left\langle \hat{\rho}%
_{b}^{\dagger}\hat{h}\right\rangle \left\langle \hat{\rho}_{a}\hat{h}%
^{\dagger}\right\rangle
\end{array}
\right)  \left(  1-e^{-i\varphi}\right)  \left(  1+e^{i\varphi}\right)  +\\
\left(
\begin{array}
[c]{c}%
\left\langle \hat{\rho}_{a}\hat{g}^{\dagger}\right\rangle \left\langle
\hat{\rho}_{a}^{\dagger}\hat{h}\right\rangle +\left\langle \hat{\rho}_{b}%
\hat{h}^{\dagger}\right\rangle \left\langle \hat{\rho}_{a}^{\dagger}\hat
{h}\right\rangle \\
-\left\langle \hat{\rho}_{a}\hat{g}^{\dagger}\right\rangle \left\langle
\hat{\rho}_{b}^{\dagger}\hat{g}\right\rangle -\left\langle \hat{\rho}_{b}%
\hat{h}^{\dagger}\right\rangle \left\langle \hat{\rho}_{b}^{\dagger}\hat
{g}\right\rangle
\end{array}
\right)  \left(  1-e^{i\varphi}\right)  \left(  1+e^{-i\varphi}\right)  ]
\end{array}
\right]  , \label{VARID}%
\end{equation}
where
\[
\hat{R}_{1}=\hat{\rho}_{a}^{\dagger}\hat{\rho}_{a}^{\dagger}\left(
1-e^{-i\varphi}\right)  ^{2}-\hat{\rho}_{b}^{\dagger}\hat{\rho}_{b}^{\dagger
}\left(  1+e^{-i\varphi}\right)  ^{2}-2i\hat{\rho}_{a}^{\dagger}\hat{\rho}%
_{b}^{\dagger}\left(  1-e^{-i\varphi}\right)  \left(  1+e^{-i\varphi}\right)
,
\]%
\[
\hat{R}_{2}=\left[
\begin{array}
[c]{c}%
\hat{\rho}_{a}\hat{\rho}_{a}^{\dagger}\left(  1-e^{i\varphi}\right)  \left(
1-e^{-i\varphi}\right)  +\hat{\rho}_{b}\hat{\rho}_{b}^{\dagger}\left(
1+e^{-i\varphi}\right)  \left(  1+e^{i\varphi}\right) \\
-i\hat{\rho}_{a}\hat{\rho}_{b}^{\dagger}\left(  1-e^{i\varphi}\right)  \left(
1+e^{-i\varphi}\right)  +i\hat{\rho}_{a}^{\dagger}\hat{\rho}_{b}\left(
1+e^{i\varphi}\right)  \left(  1-e^{-i\varphi}\right)
\end{array}
\right]  ,
\]%
\[
\hat{R}_{3}=\left[
\begin{array}
[c]{c}%
\hat{\rho}_{a}^{\dagger}\hat{\rho}_{a}\left(  1-e^{-i\varphi}\right)  \left(
1-e^{i\varphi}\right)  +\hat{\rho}_{b}^{\dagger}\hat{\rho}_{b}\left(
1+e^{-i\varphi}\right)  \left(  1+e^{i\varphi}\right) \\
-i\hat{\rho}_{a}\hat{\rho}_{b}^{\dagger}\left(  1+e^{-i\varphi}\right)
\left(  1-e^{i\varphi}\right)  +i\hat{\rho}_{a}^{\dagger}\hat{\rho}_{b}\left(
1-e^{-i\varphi}\right)  \left(  1+e^{i\varphi}\right)
\end{array}
\right]  ,
\]%
\[
\hat{R}_{4}=\hat{\rho}_{b}^{\dagger}\hat{\rho}_{b}^{\dagger}\left(
1-e^{-i\varphi}\right)  ^{2}-\hat{\rho}_{a}^{\dagger}\hat{\rho}_{a}^{\dagger
}\left(  1+e^{-i\varphi}\right)  ^{2}+2i\hat{\rho}_{a}^{\dagger}\hat{\rho}%
_{b}^{\dagger}\left(  1-e^{-i\varphi}\right)  \left(  1+e^{-i\varphi}\right)
,
\]%
\[
\hat{R}_{5}=\left[
\begin{array}
[c]{c}%
\hat{\rho}_{b}\hat{\rho}_{b}^{\dagger}\left(  1-e^{i\varphi}\right)  \left(
1-e^{-i\varphi}\right)  +\hat{\rho}_{a}\hat{\rho}_{a}^{\dagger}\left(
1+e^{i\varphi}\right)  \left(  1+e^{-i\varphi}\right) \\
+i\hat{\rho}_{a}^{\dagger}\hat{\rho}_{b}\left(  1-e^{i\varphi}\right)  \left(
1+e^{-i\varphi}\right)  -i\hat{\rho}_{a}\hat{\rho}_{b}^{\dagger}\left(
1+e^{i\varphi}\right)  \left(  1-e^{-i\varphi}\right)
\end{array}
\right]  ,
\]%
\[
\hat{R}_{6}=\left[
\begin{array}
[c]{c}%
\hat{\rho}_{b}^{\dagger}\hat{\rho}_{b}\left(  1-e^{-i\varphi}\right)  \left(
1-e^{i\varphi}\right)  +\hat{\rho}_{a}^{\dagger}\hat{\rho}_{a}\left(
1+e^{-i\varphi}\right)  \left(  1+e^{i\varphi}\right) \\
+i\hat{\rho}_{a}^{\dagger}\hat{\rho}_{b}\left(  1+e^{-i\varphi}\right)
\left(  1-e^{i\varphi}\right)  -i\hat{\rho}_{a}\hat{\rho}_{b}^{\dagger}\left(
1-e^{-i\varphi}\right)  \left(  1+e^{i\varphi}\right)
\end{array}
\right]  ,
\]%
\[
\hat{R}_{7}=2\left[
\begin{array}
[c]{c}%
\hat{\rho}_{a}^{\dagger}\hat{\rho}_{b}^{\dagger}\left(  1-e^{-i\varphi
}\right)  ^{2}+\hat{\rho}_{a}^{\dagger}\hat{\rho}_{b}^{\dagger}\left(
1+e^{-i\varphi}\right)  ^{2}\\
+i\hat{\rho}_{a}^{\dagger}\hat{\rho}_{a}^{\dagger}\left(  1-e^{-i\varphi
}\right)  \left(  1+e^{-i\varphi}\right)  -i\hat{\rho}_{b}^{\dagger}\hat{\rho
}_{b}^{\dagger}\left(  1-e^{-i\varphi}\right)  \left(  1+e^{-i\varphi}\right)
\end{array}
\right]  ,
\]%
\[
\hat{R}_{8}=\left[
\begin{array}
[c]{c}%
2\left[  \hat{\rho}_{a}\hat{\rho}_{b}^{\dagger}\left(  1-e^{i\varphi}\right)
\left(  1-e^{-i\varphi}\right)  -\hat{\rho}_{a}^{\dagger}\hat{\rho}_{b}\left(
1+e^{-i\varphi}\right)  \left(  1+e^{i\varphi}\right)  \right] \\
+i\left[  \left(  \hat{\rho}_{a}\hat{\rho}_{a}^{\dagger}+\hat{\rho}%
_{a}^{\dagger}\hat{\rho}_{a}\right)  \left(  1+e^{-i\varphi}\right)  \left(
1-e^{i\varphi}\right)  +\left(  \hat{\rho}_{b}\hat{\rho}_{b}^{\dagger}%
+\hat{\rho}_{b}^{\dagger}\hat{\rho}_{b}\right)  \left(  1-e^{-i\varphi
}\right)  \left(  1+e^{i\varphi}\right)  \right]
\end{array}
\right]  ;
\]
and
\begin{equation}
\Delta^{2}C_{\gamma,\varepsilon}=\left[
\begin{array}
[c]{c}%
\left\langle \hat{Q}_{1}\hat{e}\hat{e}\right\rangle +\left\langle \hat{Q}%
_{1}^{\dagger}\hat{e}^{\dagger}\hat{e}^{\dagger}\right\rangle +\left\langle
\hat{Q}_{2}\hat{e}^{\dagger}\hat{e}\right\rangle +\left\langle \hat{Q}_{3}%
\hat{e}\hat{e}^{\dagger}\right\rangle \\
+\left\langle \hat{Q}_{4}\hat{f}\hat{f}\right\rangle +\left\langle \hat{Q}%
_{4}^{\dagger}\hat{f}^{\dagger}\hat{f}^{\dagger}\right\rangle +\left\langle
\hat{Q}_{5}\hat{f}^{\dagger}\hat{f}\right\rangle +\left\langle \hat{Q}_{6}%
\hat{f}\hat{f}^{\dagger}\right\rangle \\
+\left\langle \hat{Q}_{7}\hat{e}\hat{f}\right\rangle +\left\langle \hat{Q}%
_{7}^{\dagger}\hat{e}^{\dagger}\hat{f}^{\dagger}\right\rangle -\left\langle
\hat{Q}_{8}\hat{e}^{\dagger}\hat{f}\right\rangle -\left\langle \hat{Q}%
_{8}^{\dagger}\hat{e}\hat{f}^{\dagger}\right\rangle \\
-\left\langle \hat{e}^{\dagger}\hat{g}\right\rangle ^{2}-\left\langle \hat
{e}\hat{g}^{\dagger}\right\rangle ^{2}-\left\langle \hat{f}^{\dagger}\hat
{h}\right\rangle ^{2}-\left\langle \hat{f}\hat{h}^{\dagger}\right\rangle
^{2}+\left\langle \hat{e}^{\dagger}\hat{h}\right\rangle ^{2}\\
+\left\langle \hat{e}\hat{h}^{\dagger}\right\rangle ^{2}+\left\langle \hat
{f}^{\dagger}\hat{g}\right\rangle ^{2}+\left\langle \hat{f}\hat{g}^{\dagger
}\right\rangle ^{2}-2(\left\langle \hat{e}^{\dagger}\hat{g}\right\rangle
\left\langle \hat{e}\hat{g}^{\dagger}\right\rangle \\
+\left\langle \hat{f}^{\dagger}\hat{h}\right\rangle \left\langle \hat{f}%
\hat{h}^{\dagger}\right\rangle +\left\langle \hat{e}^{\dagger}\hat
{h}\right\rangle \left\langle \hat{e}\hat{h}^{\dagger}\right\rangle
+\left\langle \hat{f}^{\dagger}\hat{g}\right\rangle \left\langle \hat{f}%
\hat{g}^{\dagger}\right\rangle +\left\langle \hat{e}^{\dagger}\hat
{h}\right\rangle \left\langle \hat{f}^{\dagger}\hat{g}\right\rangle \\
+\left\langle \hat{e}\hat{h}^{\dagger}\right\rangle \left\langle \hat{f}%
\hat{g}^{\dagger}\right\rangle -\left\langle \hat{e}^{\dagger}\hat
{g}\right\rangle \left\langle \hat{f}^{\dagger}\hat{h}\right\rangle
-\left\langle \hat{e}\hat{g}^{\dagger}\right\rangle \left\langle \hat{f}%
\hat{h}^{\dagger}\right\rangle -\left\langle \hat{e}^{\dagger}\hat
{g}\right\rangle \left\langle \hat{f}\hat{h}^{\dagger}\right\rangle \\
-\left\langle \hat{e}\hat{g}^{\dagger}\right\rangle \left\langle \hat
{f}^{\dagger}\hat{h}\right\rangle -\left\langle \hat{e}^{\dagger}\hat
{h}\right\rangle \left\langle \hat{f}\hat{g}^{\dagger}\right\rangle
-\left\langle \hat{e}\hat{h}^{\dagger}\right\rangle \left\langle \hat
{f}^{\dagger}\hat{g}\right\rangle )-2i(\left\langle \hat{e}^{\dagger}\hat
{g}\right\rangle \left\langle \hat{e}^{\dagger}\hat{h}\right\rangle \\
+\left\langle \hat{e}\hat{g}^{\dagger}\right\rangle \left\langle \hat
{e}^{\dagger}\hat{h}\right\rangle +\left\langle \hat{e}\hat{g}^{\dagger
}\right\rangle \left\langle \hat{f}\hat{g}^{\dagger}\right\rangle
+\left\langle \hat{e}^{\dagger}\hat{g}\right\rangle \left\langle \hat{f}%
\hat{g}^{\dagger}\right\rangle +\left\langle \hat{f}\hat{h}^{\dagger
}\right\rangle \left\langle \hat{e}\hat{h}^{\dagger}\right\rangle \\
+\left\langle \hat{e}\hat{h}^{\dagger}\right\rangle \left\langle \hat
{f}^{\dagger}\hat{h}\right\rangle +\left\langle \hat{f}^{\dagger}\hat
{h}\right\rangle \left\langle \hat{f}^{\dagger}\hat{g}\right\rangle
+\left\langle \hat{f}\hat{h}^{\dagger}\right\rangle \left\langle \hat
{f}^{\dagger}\hat{g}\right\rangle -\left\langle \hat{e}\hat{g}^{\dagger
}\right\rangle \left\langle \hat{e}\hat{h}^{\dagger}\right\rangle \\
-\left\langle \hat{e}^{\dagger}\hat{g}\right\rangle \left\langle \hat{e}%
\hat{h}^{\dagger}\right\rangle -\left\langle \hat{e}^{\dagger}\hat
{g}\right\rangle \left\langle \hat{f}^{\dagger}\hat{g}\right\rangle
-\left\langle \hat{e}\hat{g}^{\dagger}\right\rangle \left\langle \hat
{f}^{\dagger}\hat{g}\right\rangle -\left\langle \hat{f}^{\dagger}\hat
{h}\right\rangle \left\langle \hat{e}^{\dagger}\hat{h}\right\rangle \\
-\left\langle \hat{f}\hat{h}^{\dagger}\right\rangle \left\langle \hat
{e}^{\dagger}\hat{h}\right\rangle -\left\langle \hat{f}\hat{h}^{\dagger
}\right\rangle \left\langle \hat{f}\hat{g}^{\dagger}\right\rangle
-\left\langle \hat{f}^{\dagger}\hat{h}\right\rangle \left\langle \hat{f}%
\hat{g}^{\dagger}\right\rangle )
\end{array}
\right]  ,\nonumber
\end{equation}

where
\[
\hat{Q}_{1}=\hat{g}^{\dagger}\hat{g}^{\dagger}-\hat{h}^{\dagger}\hat
{h}^{\dagger}-2i\hat{g}^{\dagger}\hat{h}^{\dagger},
\]
\[
\hat{Q}_{2}=\hat{g}\hat{g}^{\dagger}+\hat{h}\hat{h}^{\dagger}+i\left(  \hat
{g}^{\dagger}\hat{h}-\hat{g}\hat{h}^{\dagger}\right)  ,
\]
\[
\hat{Q}_{3}=\hat{g}^{\dagger}\hat{g}+\hat{h}^{\dagger}\hat{h}+i\left(  \hat
{g}^{\dagger}\hat{h}-\hat{g}\hat{h}^{\dagger}\right)  ,
\]
\[
\hat{Q}_{4}=\hat{h}^{\dagger}\hat{h}^{\dagger}-\hat{g}^{\dagger}\hat
{g}^{\dagger}-2i\hat{g}^{\dagger}\hat{h}^{\dagger},
\]
\[
\hat{Q}_{5}=\hat{g}\hat{g}^{\dagger}+\hat{h}\hat{h}^{\dagger}+i\left(  \hat
{g}\hat{h}^{\dagger}-\hat{g}^{\dagger}\hat{h}\right)  ,
\]
\[
\hat{Q}_{6}=\hat{g}^{\dagger}\hat{g}+\hat{h}^{\dagger}\hat{h}+i\left(  \hat
{g}\hat{h}^{\dagger}-\hat{g}^{\dagger}\hat{h}\right)  ,
\]
\[
\hat{Q}_{7}=2i\left(  \hat{g}^{\dagger}\hat{g}^{\dagger}+\hat{h}^{\dagger}%
\hat{h}^{\dagger}\right)  ,
\]
and
\[
\hat{Q}_{8}=2\left(  \hat{g}\hat{h}^{\dagger}+\hat{g}^{\dagger}\hat{h}\right)
-i\left(  \hat{g}\hat{g}^{\dagger}+\hat{g}^{\dagger}\hat{g}\right)  +i\left(
\hat{h}\hat{h}^{\dagger}+\hat{h}^{\dagger}\hat{h}\right)  .
\]

\section{MEASUREMENT\ OPERATOR\ CROSS\ CORRELATIONS}

The expressions for the measurement operator cross correlations that are
required to evaluate the numerator in eq.(\ref{D2PHI}) are
\[
\left\langle \hat{D}_{\alpha;\alpha,\gamma}\right\rangle =8\left[
\begin{array}
[c]{c}%
\{\left\langle \hat{S}_{1}\right\rangle +\left\langle \hat{S}_{1}^{\dagger
}\right\rangle -\left\langle \hat{S}_{2}\right\rangle -\left\langle \hat
{S}_{2}^{\dagger}\right\rangle \}\cos\varphi-\\
\{\left\langle \hat{S}_{3}\right\rangle +\left\langle \hat{S}_{3}^{\dagger
}\right\rangle -\left\langle \hat{S}_{4}\right\rangle -\left\langle \hat
{S}_{4}^{\dagger}\right\rangle \}\sin\varphi-\\
2\{\left(  \left\langle \hat{\rho}_{b}^{\dagger}\hat{\rho}_{b}\right\rangle
-\left\langle \hat{\rho}_{a}^{\dagger}\hat{\rho}_{a}\right\rangle \right)
\cos\varphi-\left(  \left\langle \hat{\rho}_{a}^{\dagger}\hat{\rho}%
_{b}\right\rangle +\left\langle \hat{\rho}_{a}\hat{\rho}_{b}^{\dagger
}\right\rangle \right)  \sin\varphi\}\cdot\\
\{\left(  \left\langle \hat{\rho}_{b}^{\dagger}\hat{f}\right\rangle
+i\left\langle \hat{\rho}_{a}^{\dagger}\hat{f}\right\rangle \right)  +\left(
\left\langle \hat{\rho}_{b}\hat{f}^{\dagger}\right\rangle -i\left\langle
\hat{\rho}_{a}\hat{f}^{\dagger}\right\rangle \right)  -\\
\left(  \left\langle \hat{\rho}_{a}^{\dagger}\hat{e}\right\rangle
+i\left\langle \hat{\rho}_{b}^{\dagger}e\right\rangle \right)  e^{-i\varphi
}-\left(  \left\langle \hat{\rho}_{a}\hat{e}^{\dagger}\right\rangle
-i\left\langle \hat{\rho}_{b}\hat{e}^{\dagger}\right\rangle \right)
e^{i\varphi}\}
\end{array}
\right]  ,
\]
where
\[
\hat{S}_{1}=\left[  \left(  \hat{\rho}_{b}^{\dagger}\hat{\rho}_{b}\hat{\rho
}_{b}^{\dagger}+\hat{\rho}_{b}^{\dagger}\hat{\rho}_{b}^{\dagger}\hat{\rho}%
_{b}-2\hat{\rho}_{a}^{\dagger}\hat{\rho}_{a}\hat{\rho}_{b}^{\dagger}\right)
+i\left(  2\hat{\rho}_{a}^{\dagger}\hat{\rho}_{b}^{\dagger}\hat{\rho}_{b}%
-\hat{\rho}_{a}^{\dagger}\hat{\rho}_{a}\hat{\rho}_{a}^{\dagger}-\hat{\rho}%
_{a}^{\dagger}\hat{\rho}_{a}^{\dagger}\hat{\rho}_{a}\right)  \right]  \hat
{f},
\]%
\[
\hat{S}_{2}=\left[  \left(  2\hat{\rho}_{a}^{\dagger}\hat{\rho}_{b}^{\dagger
}\hat{\rho}_{b}-\hat{\rho}_{a}^{\dagger}\hat{\rho}_{a}\hat{\rho}_{a}^{\dagger
}-\hat{\rho}_{a}^{\dagger}\hat{\rho}_{a}^{\dagger}\hat{\rho}_{a}\right)
-i\left(  2\hat{\rho}_{a}^{\dagger}\hat{\rho}_{a}\hat{\rho}_{b}^{\dagger}%
-\hat{\rho}_{b}^{\dagger}\hat{\rho}_{b}\hat{\rho}_{b}^{\dagger}-\hat{\rho}%
_{b}^{\dagger}\hat{\rho}_{b}^{\dagger}\hat{\rho}_{b}\right)  \right]  \hat
{e}e^{-i\varphi},
\]%
\[
\hat{S}_{3}=\left[  \left(  2\hat{\rho}_{a}\hat{\rho}_{b}^{\dagger}\hat{\rho
}_{b}^{\dagger}+\hat{\rho}_{a}^{\dagger}\hat{\rho}_{b}\hat{\rho}_{b}^{\dagger
}+\hat{\rho}_{a}^{\dagger}\hat{\rho}_{b}^{\dagger}\hat{\rho}_{b}\right)
+i\left(  2\hat{\rho}_{a}^{\dagger}\hat{\rho}_{a}^{\dagger}\hat{\rho}_{b}%
+\hat{\rho}_{a}\hat{\rho}_{a}^{\dagger}\hat{\rho}_{b}^{\dagger}+\hat{\rho}%
_{a}^{\dagger}\hat{\rho}_{a}\hat{\rho}_{b}^{\dagger}\right)  \right]  \hat
{f},
\]%
\[
\hat{S}_{4}=\left[  \left(  2\hat{\rho}_{a}^{\dagger}\hat{\rho}_{a}^{\dagger
}\hat{\rho}_{b}+\hat{\rho}_{a}\hat{\rho}_{a}^{\dagger}\hat{\rho}_{b}^{\dagger
}+\hat{\rho}_{a}^{\dagger}\hat{\rho}_{a}\hat{\rho}_{b}^{\dagger}\right)
+i\left(  2\hat{\rho}_{a}\hat{\rho}_{b}^{\dagger}\hat{\rho}_{b}^{\dagger}%
+\hat{\rho}_{a}^{\dagger}\hat{\rho}_{b}\hat{\rho}_{b}^{\dagger}+\hat{\rho}%
_{a}^{\dagger}\hat{\rho}_{b}^{\dagger}\hat{\rho}_{b}\right)  \right]  \hat
{e}e^{-i\varphi};
\]%
\[
\left\langle \hat{D}_{\alpha;\alpha,\varepsilon}\right\rangle =4\left[
\begin{array}
[c]{c}%
\{\left\langle \hat{U}_{1}\right\rangle +\left\langle \hat{U}_{1}^{\dagger
}\right\rangle -\left\langle \hat{U}_{2}\right\rangle -\left\langle \hat
{U}_{2}^{\dagger}\right\rangle \}\cos\varphi-\\
\{\left\langle \hat{U}_{3}\right\rangle +\left\langle \hat{U}_{3}^{\dagger
}\right\rangle +\left\langle \hat{U}_{4}\right\rangle +\left\langle \hat
{U}_{4}^{\dagger}\right\rangle \}\sin\varphi-\\
2\{\left(  \left\langle \hat{\rho}_{b}^{\dagger}\hat{\rho}_{b}\right\rangle
-\left\langle \hat{\rho}_{a}^{\dagger}\hat{\rho}_{a}\right\rangle \right)
\cos\varphi-\left(  \left\langle \hat{\rho}_{a}^{\dagger}\hat{\rho}%
_{b}\right\rangle +\left\langle \hat{\rho}_{a}\hat{\rho}_{b}^{\dagger
}\right\rangle \right)  \sin\varphi\}\cdot\\
\{\left(  \left\langle \hat{\rho}_{a}^{\dagger}\hat{g}\right\rangle
+\left\langle \hat{\rho}_{b}^{\dagger}\hat{h}\right\rangle \right)  \left(
1-e^{-i\varphi}\right)  +\left(  \left\langle \hat{\rho}_{a}\hat{g}^{\dagger
}\right\rangle +\left\langle \hat{\rho}_{b}\hat{h}^{\dagger}\right\rangle
\right)  \left(  1-e^{i\varphi}\right)  +\\
i\left(  \left\langle \hat{\rho}_{a}^{\dagger}\hat{h}\right\rangle
-\left\langle \hat{\rho}_{b}^{\dagger}\hat{g}\right\rangle \right)  \left(
1+e^{-i\varphi}\right)  +i\left(  \left\langle \hat{\rho}_{b}\hat{g}^{\dagger
}\right\rangle -\left\langle \hat{\rho}_{a}\hat{h}^{\dagger}\right\rangle
\right)  \left(  1+e^{i\varphi}\right)  \}
\end{array}
\right]  ,
\]
where
\[
\hat{U}_{1}=\left(  2\hat{\rho}_{a}^{\dagger}\hat{\rho}_{b}^{\dagger}\hat
{\rho}_{b}-\hat{\rho}_{a}^{\dagger}\hat{\rho}_{a}\hat{\rho}_{a}^{\dagger}%
-\hat{\rho}_{a}^{\dagger}\hat{\rho}_{a}^{\dagger}\hat{\rho}_{a}\right)
\left[  \hat{g}\left(  1-e^{-i\varphi}\right)  +i\hat{h}\left(  1+e^{-i\varphi
}\right)  \right]  ,
\]%
\[
\hat{U}_{2}=\left(  2\hat{\rho}_{a}^{\dagger}\hat{\rho}_{a}\hat{\rho}%
_{b}^{\dagger}-\hat{\rho}_{b}^{\dagger}\hat{\rho}_{b}\hat{\rho}_{b}^{\dagger
}-\hat{\rho}_{b}^{\dagger}\hat{\rho}_{b}^{\dagger}\hat{\rho}_{b}\right)
\left[  \hat{h}\left(  1-e^{-i\varphi}\right)  -i\hat{g}\left(  1+e^{-i\varphi
}\right)  \right]  ,
\]%
\[
\hat{U}_{3}=\left(  2\hat{\rho}_{a}^{\dagger}\hat{\rho}_{a}^{\dagger}\hat
{\rho}_{b}+\hat{\rho}_{a}\hat{\rho}_{a}^{\dagger}\hat{\rho}_{b}^{\dagger}%
+\hat{\rho}_{a}^{\dagger}\hat{\rho}_{a}\hat{\rho}_{b}^{\dagger}\right)
\left[  \hat{g}\left(  1-e^{-i\varphi}\right)  +i\hat{h}\left(  1+e^{-i\varphi
}\right)  \right]  ,
\]%
\[
\hat{U}_{4}=\left(  2\hat{\rho}_{a}\hat{\rho}_{b}^{\dagger}\hat{\rho}%
_{b}^{\dagger}+\hat{\rho}_{a}^{\dagger}\hat{\rho}_{b}\hat{\rho}_{b}^{\dagger
}+\hat{\rho}_{a}^{\dagger}\hat{\rho}_{b}^{\dagger}\hat{\rho}_{b}\right)
\left[  \hat{h}\left(  1-e^{-i\varphi}\right)  -i\hat{g}\left(  1+e^{-i\varphi
}\right)  \right]  ;
\]%
\[
\left\langle \hat{D}_{\gamma;\alpha,\gamma}\right\rangle =4\left[
\begin{array}
[c]{c}%
\left\langle \hat{V}_{1}\right\rangle +\left\langle \hat{V}_{1}^{\dagger
}\right\rangle +\left\langle \hat{V}_{2}\right\rangle +\left\langle \hat
{V}_{2}^{\dagger}\right\rangle -\left\langle \hat{V}_{3}\right\rangle
-\left\langle \hat{V}_{3}^{\dagger}\right\rangle -\left\langle \hat{V}%
_{4}\right\rangle -\left\langle \hat{V}_{4}^{\dagger}\right\rangle -\\
2\left(  \left\langle \hat{e}^{\dagger}\hat{f}\right\rangle -\left\langle
\hat{e}\hat{f}^{\dagger}\right\rangle \right)  [\left(  \left\langle \hat
{\rho}_{a}\hat{f}^{\dagger}\right\rangle +i\left\langle \hat{\rho}_{b}\hat
{f}^{\dagger}\right\rangle \right)  -\left(  \left\langle \hat{\rho}%
_{a}^{\dagger}\hat{f}\right\rangle -i\left\langle \hat{\rho}_{b}^{\dagger}%
\hat{f}\right\rangle \right)  +\\
\left(  \left\langle \hat{\rho}_{b}^{\dagger}\hat{e}\right\rangle
-i\left\langle \hat{\rho}_{a}^{\dagger}\hat{e}\right\rangle \right)
e^{-i\varphi}-\left(  \left\langle \hat{\rho}_{b}\hat{e}^{\dagger
}\right\rangle +i\left\langle \hat{\rho}_{a}\hat{e}^{\dagger}\right\rangle
\right)  e^{i\varphi}]
\end{array}
\right]  ,
\]
where
\[
\hat{V}_{1}=\left(  \hat{\rho}_{a}+i\hat{\rho}_{b}\right)  \hat{e}^{\dagger
}\hat{f}\hat{f}^{\dagger},
\]%
\[
\hat{V}_{2}=\left(  \hat{\rho}_{a}+i\hat{\rho}_{b}\right)  \hat{e}^{\dagger
}\hat{f}^{\dagger}\hat{f},
\]%
\[
\hat{V}_{3}=2\left(  \hat{\rho}_{a}+i\hat{\rho}_{b}\right)  \hat{e}\hat
{f}^{\dagger}\hat{f}^{\dagger},
\]%
\[
\hat{V}_{4}=\left(  \hat{\rho}_{b}+i\hat{\rho}_{a}\right)  \left(  2\hat
{e}^{\dagger}\hat{e}^{\dagger}\hat{f}-\hat{e}^{\dagger}\hat{e}\hat{f}%
^{\dagger}-\hat{e}\hat{e}^{\dagger}\hat{f}^{\dagger}\right)  e^{i\varphi};
\]%
\[
\left\langle \hat{D}_{\gamma;\gamma,\varepsilon}\right\rangle =2\left[
\begin{array}
[c]{c}%
\left\langle \hat{W}_{1}\right\rangle +\left\langle \hat{W}_{1}^{\dagger
}\right\rangle +\left\langle \hat{W}_{2}\right\rangle +\left\langle \hat
{W}_{2}^{\dagger}\right\rangle -2\left(  \left\langle \hat{e}^{\dagger}\hat
{f}\right\rangle -\left\langle \hat{e}\hat{f}^{\dagger}\right\rangle \right)
\cdot\\
\lbrack\left(  \left\langle \hat{f}^{\dagger}\hat{g}\right\rangle
-i\left\langle \hat{f}^{\dagger}\hat{h}\right\rangle \right)  -\left(
\left\langle \hat{f}\hat{g}^{\dagger}\right\rangle +i\left\langle \hat{f}%
\hat{h}^{\dagger}\right\rangle \right)  -\\
\left(  \left\langle \hat{e}^{\dagger}\hat{h}\right\rangle -i\left\langle
\hat{e}^{\dagger}\hat{g}\right\rangle \right)  +\left(  \left\langle \hat
{e}\hat{h}^{\dagger}\right\rangle +i\left\langle \hat{e}\hat{g}^{\dagger
}\right\rangle \right)  ]
\end{array}
\right]  ,
\]
where
\[
\hat{W}_{1}=\left(  \hat{e}^{\dagger}\hat{f}\hat{f}^{\dagger}+\hat{e}%
^{\dagger}\hat{f}^{\dagger}\hat{f}-2\hat{e}\hat{f}^{\dagger}\hat{f}^{\dagger
}\right)  \left(  \hat{g}-i\hat{h}\right)  ,
\]%
\[
\hat{W}_{2}=\left(  \hat{e}^{\dagger}\hat{e}\hat{f}^{\dagger}+\hat{e}\hat
{e}^{\dagger}\hat{f}^{\dagger}-2\hat{e}^{\dagger}\hat{e}^{\dagger}\hat
{f}\right)  \left(  \hat{h}-i\hat{g}\right)  ;
\]%
\begin{equation}
\left\langle \hat{D}_{\varepsilon;\alpha,\varepsilon}\right\rangle =\left[
\begin{array}
[c]{c}%
\left\langle \hat{H}_{1}\right\rangle +\left\langle \hat{H}_{1}^{\dagger
}\right\rangle +\left\langle \hat{H}_{2}\right\rangle +\left\langle \hat
{H}_{2}^{\dagger}\right\rangle +\left\langle \hat{H}_{3}\right\rangle
+\left\langle \hat{H}_{3}^{\dagger}\right\rangle +\left\langle \hat{H}%
_{4}\right\rangle +\left\langle \hat{H}_{4}^{\dagger}\right\rangle \\
-2\left[  \left\langle \hat{g}^{\dagger}\hat{g}\right\rangle -\left\langle
\hat{h}^{\dagger}\hat{h}\right\rangle \right]  \cdot\lbrack\left(
\left\langle \hat{\rho}_{a}^{\dagger}\hat{g}\right\rangle +\left\langle
\hat{\rho}_{b}^{\dagger}\hat{h}\right\rangle \right)  \left(  1-e^{-i\varphi
}\right) \\
+\left(  \left\langle \hat{\rho}_{a}\hat{g}^{\dagger}\right\rangle
+\left\langle \hat{\rho}_{b}\hat{h}^{\dagger}\right\rangle \right)  \left(
1-e^{i\varphi}\right)  +i\left(  \left\langle \hat{\rho}_{a}^{\dagger}\hat
{h}\right\rangle -\left\langle \hat{\rho}_{b}^{\dagger}\hat{g}\right\rangle
\right)  \left(  1+e^{-i\varphi}\right) \\
-i\left(  \left\langle \hat{\rho}_{a}\hat{h}^{\dagger}\right\rangle
-\left\langle \hat{\rho}_{b}\hat{g}^{\dagger}\right\rangle \right)  \left(
1+e^{i\varphi}\right)  ]
\end{array}
\right]  ,\nonumber
\end{equation}
where
\[
\hat{H}_{1}=\hat{\rho}_{b}\left(  2\hat{g}^{\dagger}\hat{g}\hat{h}^{\dagger
}-\hat{h}^{\dagger}\hat{h}\hat{h}^{\dagger}-\hat{h}^{\dagger}\hat{h}^{\dagger
}\hat{h}\right)  \left(  1-e^{i\varphi}\right)  ,
\]%
\[
\hat{H}_{2}=-\hat{\rho}_{a}\left(  2\hat{g}^{\dagger}\hat{h}^{\dagger}\hat
{h}-\hat{g}^{\dagger}\hat{g}\hat{g}^{\dagger}-\hat{g}^{\dagger}\hat
{g}^{\dagger}\hat{g}\right)  \left(  1-e^{i\varphi}\right)  ,
\]%
\[
\hat{H}_{3}=-i\hat{\rho}_{a}\left(  2\hat{g}^{\dagger}\hat{g}\hat{h}^{\dagger
}-\hat{h}^{\dagger}\hat{h}\hat{h}^{\dagger}-\hat{h}^{\dagger}\hat{h}^{\dagger
}\hat{h}\right)  \left(  1+e^{i\varphi}\right)  ,
\]%
\[
\hat{H}_{4}=-i\hat{\rho}_{b}\left(  2\hat{g}^{\dagger}\hat{h}^{\dagger}\hat
{h}-\hat{g}^{\dagger}\hat{g}\hat{g}^{\dagger}-\hat{g}^{\dagger}\hat
{g}^{\dagger}\hat{g}\right)  \left(  1+e^{i\varphi}\right)  ;
\]%
\begin{equation}
\left\langle \hat{D}_{\varepsilon;\gamma,\varepsilon}\right\rangle =\left[
\begin{array}
[c]{c}%
\left\langle \hat{L}_{1}\right\rangle +\left\langle \hat{L}_{1}^{\dagger
}\right\rangle +\left\langle \hat{L}_{2}\right\rangle +\left\langle \hat
{L}_{2}^{\dagger}\right\rangle +\left\langle \hat{L}_{3}\right\rangle
+\left\langle \hat{L}_{3}^{\dagger}\right\rangle +\left\langle \hat{L}%
_{4}\right\rangle +\left\langle \hat{L}_{4}^{\dagger}\right\rangle \\
-2\left(  \left\langle \hat{g}^{\dagger}\hat{g}\right\rangle -\left\langle
\hat{h}^{\dagger}\hat{h}\right\rangle \right)  [\left(  \left\langle \hat
{e}^{\dagger}\hat{g}\right\rangle +\left\langle \hat{e}\hat{g}^{\dagger
}\right\rangle \right)  -\left(  \left\langle \hat{f}^{\dagger}\hat
{h}\right\rangle +\left\langle \hat{f}\hat{h}^{\dagger}\right\rangle \right)
\\
+i\left(  \left\langle \hat{e}^{\dagger}\hat{h}\right\rangle -\left\langle
\hat{e}\hat{h}^{\dagger}\right\rangle \right)  -i\left(  \left\langle \hat
{f}^{\dagger}\hat{g}\right\rangle -\left\langle \hat{f}\hat{g}^{\dagger
}\right\rangle \right)  ]
\end{array}
\right]  ,\nonumber
\end{equation}
where
\[
\hat{L}_{1}=-\hat{e}\left(  2\hat{g}^{\dagger}\hat{h}^{\dagger}\hat{h}-\hat
{g}^{\dagger}\hat{g}\hat{g}^{\dagger}-\hat{g}^{\dagger}\hat{g}^{\dagger}%
\hat{g}\right)  ,
\]%
\[
\hat{L}_{2}=-\hat{f}\left(  2\hat{g}^{\dagger}\hat{g}\hat{h}^{\dagger}-\hat
{h}^{\dagger}\hat{h}\hat{h}^{\dagger}-\hat{h}^{\dagger}\hat{h}^{\dagger}%
\hat{h}\right)  ,
\]%
\[
\hat{L}_{3}=-i\hat{e}\left(  2\hat{g}^{\dagger}\hat{g}\hat{h}^{\dagger}%
-\hat{h}^{\dagger}\hat{h}\hat{h}^{\dagger}-\hat{h}^{\dagger}\hat{h}^{\dagger
}\hat{h}\right)  ,
\]%
\[
\hat{L}_{4}=-i\hat{f}\left(  2\hat{g}^{\dagger}\hat{h}^{\dagger}\hat{h}%
-\hat{g}^{\dagger}\hat{g}\hat{g}^{\dagger}-\hat{g}^{\dagger}\hat{g}^{\dagger
}\hat{g}\right)  ;
\]%
\[
\left\langle \hat{D}_{\alpha,\gamma;\alpha,\varepsilon}\right\rangle =2\left[
\begin{array}
[c]{c}%
\left\langle \hat{K}_{1}\right\rangle +\left\langle \hat{K}_{1}^{\dagger
}\right\rangle +\left\langle \hat{K}_{2}\right\rangle +\left\langle \hat
{K}_{2}^{\dagger}\right\rangle -\left\langle \hat{K}_{3}\right\rangle
-\left\langle \hat{K}_{3}^{\dagger}\right\rangle -\\
\left\langle \hat{K}_{4}\right\rangle -\left\langle \hat{K}_{4}^{\dagger
}\right\rangle -\left\langle \hat{K}_{5}\right\rangle -\left\langle \hat
{K}_{5}^{\dagger}\right\rangle -\left\langle \hat{K}_{6}\right\rangle
-\left\langle \hat{K}_{6}^{\dagger}\right\rangle -\\
2[\left\langle \hat{\rho}_{b}^{\dagger}\hat{f}\right\rangle +\left\langle
\hat{\rho}_{b}\hat{f}^{\dagger}\right\rangle -\left\langle \hat{\rho}%
_{a}^{\dagger}\hat{e}\right\rangle e^{-i\varphi}-\left\langle \hat{\rho}%
_{a}\hat{e}^{\dagger}\right\rangle e^{i\varphi}+\\
i\left\langle \hat{\rho}_{a}^{\dagger}\hat{f}\right\rangle -i\left\langle
\hat{\rho}_{a}\hat{f}^{\dagger}\right\rangle -i\left\langle \hat{\rho}%
_{b}^{\dagger}\hat{e}\right\rangle e^{-i\varphi}+i\left\langle \hat{\rho}%
_{b}\hat{e}^{\dagger}\right\rangle e^{i\varphi}]\cdot\\
\lbrack\left(  \left\langle \hat{\rho}_{a}^{\dagger}\hat{g}\right\rangle
+\left\langle \hat{\rho}_{b}^{\dagger}\hat{h}\right\rangle \right)  \left(
1-e^{-i\varphi}\right)  +\left(  \left\langle \hat{\rho}_{a}\hat{g}^{\dagger
}\right\rangle +\left\langle \hat{\rho}_{b}\hat{h}^{\dagger}\right\rangle
\right)  \left(  1-e^{i\varphi}\right)  +\\
i\left(  \left\langle \hat{\rho}_{a}^{\dagger}\hat{h}\right\rangle
-\left\langle \hat{\rho}_{b}^{\dagger}\hat{g}\right\rangle \right)  \left(
1+e^{-i\varphi}\right)  +i\left(  \left\langle \hat{\rho}_{b}\hat{g}^{\dagger
}\right\rangle -\left\langle \hat{\rho}_{a}\hat{h}^{\dagger}\right\rangle
\right)  \left(  1+e^{i\varphi}\right)  ]
\end{array}
\right]  ,
\]
where
\[
\hat{K}_{1}=\left[
\begin{array}
[c]{c}%
2\hat{\rho}_{b}^{\dagger}\hat{\rho}_{b}^{\dagger}\hat{f}\hat{h}+2\hat{\rho
}_{a}^{\dagger}\hat{\rho}_{b}^{\dagger}\hat{f}\hat{g}+2\hat{\rho}_{a}%
^{\dagger}\hat{\rho}_{b}\hat{f}^{\dagger}\hat{g}+\hat{\rho}_{b}\hat{\rho}%
_{b}^{\dagger}\hat{f}^{\dagger}\hat{h}+\hat{\rho}_{b}^{\dagger}\hat{\rho}%
_{b}\hat{f}^{\dagger}\hat{h}+\\
i(2\hat{\rho}_{a}^{\dagger}\hat{\rho}_{a}^{\dagger}\hat{f}\hat{g}-\hat{\rho
}_{a}\hat{\rho}_{a}^{\dagger}\hat{f}^{\dagger}\hat{g}-\hat{\rho}_{a}^{\dagger
}\hat{\rho}_{a}\hat{f}^{\dagger}\hat{g}+2\hat{\rho}_{a}^{\dagger}\hat{\rho
}_{b}^{\dagger}\hat{f}\hat{h}-2\hat{\rho}_{a}\hat{\rho}_{b}^{\dagger}\hat
{f}^{\dagger}\hat{h})
\end{array}
\right]  \left(  1-e^{-i\varphi}\right)  ,
\]%
\[
\hat{K}_{2}=2\left[  \hat{\rho}_{a}^{\dagger}\hat{\rho}_{b}^{\dagger}\hat
{e}\hat{h}-\hat{\rho}_{b}^{\dagger}\hat{\rho}_{b}^{\dagger}\hat{e}\hat
{g}-i(\hat{\rho}_{a}^{\dagger}\hat{\rho}_{a}^{\dagger}\hat{e}\hat{h}-\hat
{\rho}_{a}^{\dagger}\hat{\rho}_{b}^{\dagger}\hat{e}\hat{g})\right]  \left(
1+e^{-i\varphi}\right)  e^{-i\varphi},
\]%
\[
\hat{K}_{3}=\left[
\begin{array}
[c]{c}%
2\hat{\rho}_{a}^{\dagger}\hat{\rho}_{a}^{\dagger}\hat{f}\hat{h}+\hat{\rho}%
_{a}\hat{\rho}_{a}^{\dagger}\hat{f}^{\dagger}\hat{h}+\hat{\rho}_{a}^{\dagger
}\hat{\rho}_{a}\hat{f}^{\dagger}\hat{h}-2\hat{\rho}_{a}^{\dagger}\hat{\rho
}_{b}^{\dagger}\hat{f}\hat{g}+2\hat{\rho}_{a}\hat{\rho}_{b}^{\dagger}\hat
{f}^{\dagger}\hat{g}-\\
i(2\hat{\rho}_{a}^{\dagger}\hat{\rho}_{b}^{\dagger}\hat{f}\hat{h}+2\hat{\rho
}_{a}^{\dagger}\hat{\rho}_{b}\hat{f}^{\dagger}\hat{h}-2\hat{\rho}_{b}%
^{\dagger}\hat{\rho}_{b}^{\dagger}\hat{f}\hat{g}-\hat{\rho}_{b}\hat{\rho}%
_{b}^{\dagger}\hat{f}^{\dagger}\hat{g}-\hat{\rho}_{b}^{\dagger}\hat{\rho}%
_{b}\hat{f}^{\dagger}\hat{g})
\end{array}
\right]  \left(  1+e^{-i\varphi}\right)  ,
\]%
\[
\hat{K}_{4}=2\left[  \hat{\rho}_{a}\hat{\rho}_{a}\hat{e}^{\dagger}\hat
{g}^{\dagger}+\hat{\rho}_{a}\hat{\rho}_{b}\hat{e}^{\dagger}\hat{h}^{\dagger
}-i(\hat{\rho}_{a}\hat{\rho}_{b}\hat{e}^{\dagger}\hat{g}^{\dagger}+\hat{\rho
}_{b}\hat{\rho}_{b}\hat{e}^{\dagger}\hat{h}^{\dagger})\right]  \left(
1-e^{i\varphi}\right)  e^{i\varphi},
\]%
\[
\hat{K}_{5}=\left[  2\hat{\rho}_{a}\hat{\rho}_{b}^{\dagger}\hat{e}\hat
{h}^{\dagger}-\hat{\rho}_{b}^{\dagger}\hat{\rho}_{b}\hat{e}\hat{g}^{\dagger
}-\hat{\rho}_{b}\hat{\rho}_{b}^{\dagger}\hat{e}\hat{g}^{\dagger}-i(\hat{\rho
}_{a}^{\dagger}\hat{\rho}_{a}\hat{e}\hat{h}^{\dagger}+\hat{\rho}_{a}\hat{\rho
}_{a}^{\dagger}\hat{e}\hat{h}^{\dagger}-2\hat{\rho}_{a}^{\dagger}\hat{\rho
}_{b}\hat{e}\hat{g}^{\dagger})\right]  \left(  1+e^{i\varphi}\right)
e^{-i\varphi},
\]%
\[
\hat{K}_{6}=\left[  2\hat{\rho}_{a}^{\dagger}\hat{\rho}_{b}\hat{e}\hat
{h}^{\dagger}+\hat{\rho}_{a}^{\dagger}\hat{\rho}_{a}\hat{e}\hat{g}^{\dagger
}+\hat{\rho}_{a}\hat{\rho}_{a}^{\dagger}\hat{e}\hat{g}^{\dagger}+i(2\hat{\rho
}_{a}\hat{\rho}_{b}^{\dagger}\hat{e}\hat{g}^{\dagger}+\hat{\rho}_{b}^{\dagger
}\hat{\rho}_{b}\hat{e}\hat{h}^{\dagger}+\hat{\rho}_{b}\hat{\rho}_{b}^{\dagger
}\hat{e}\hat{h}^{\dagger})\right]  \left(  1-e^{i\varphi}\right)
e^{-i\varphi};
\]%
\[
\left\langle \hat{D}_{\alpha,\gamma;\gamma,\varepsilon}\right\rangle =2\left[
\begin{array}
[c]{c}%
\left\langle \hat{O}_{1}\right\rangle +\left\langle \hat{O}_{1}^{\dagger
}\right\rangle -\left\langle \hat{O}_{2}\right\rangle -\left\langle \hat
{O}_{2}^{\dagger}\right\rangle -\left\langle \hat{O}_{3}\right\rangle
-\left\langle \hat{O}_{3}^{\dagger}\right\rangle -\\
\left\langle \hat{O}_{4}\right\rangle -\left\langle \hat{O}_{4}^{\dagger
}\right\rangle -\left\langle \hat{O}_{5}\right\rangle -\left\langle \hat
{O}_{5}^{\dagger}\right\rangle -\left\langle \hat{O}_{6}\right\rangle
-\left\langle \hat{O}_{6}^{\dagger}\right\rangle -\\
2\{\left\langle \hat{\rho}_{b}^{\dagger}\hat{f}\right\rangle +\left\langle
\hat{\rho}_{b}\hat{f}^{\dagger}\right\rangle -\left\langle \hat{\rho}%
_{a}^{\dagger}\hat{e}\right\rangle e^{-i\varphi}-\left\langle \hat{\rho}%
_{a}\hat{e}^{\dagger}\right\rangle e^{i\varphi}+\\
i\left[  \left\langle \hat{\rho}_{a}^{\dagger}\hat{f}\right\rangle
-\left\langle \hat{\rho}_{a}\hat{f}^{\dagger}\right\rangle -\left\langle
\hat{\rho}_{b}^{\dagger}\hat{e}\right\rangle e^{-i\varphi}+\left\langle
\hat{\rho}_{b}\hat{e}^{\dagger}\right\rangle e^{i\varphi}\right]  \}\cdot\\
\{\left\langle \hat{e}^{\dagger}\hat{g}\right\rangle +\left\langle \hat{e}%
\hat{g}^{\dagger}\right\rangle -\left\langle \hat{f}^{\dagger}\hat
{h}\right\rangle -\left\langle \hat{f}\hat{h}^{\dagger}\right\rangle
+i[\left\langle \hat{e}^{\dagger}\hat{h}\right\rangle -\left\langle \hat
{e}\hat{h}^{\dagger}\right\rangle -\left\langle \hat{f}^{\dagger}\hat
{g}\right\rangle +\left\langle \hat{f}\hat{g}^{\dagger}\right\rangle ]\}
\end{array}
\right]  ,
\]
where
\[
\hat{O}_{1}=2\left\{
\begin{array}
[c]{c}%
\hat{\rho}_{b}^{\dagger}\left[  \hat{e}^{\dagger}\hat{f}\hat{g}+\hat{e}\hat
{f}\hat{g}^{\dagger}-\hat{e}\hat{f}^{\dagger}\hat{g}e^{-i\varphi}+\hat{e}%
\hat{f}\hat{g}^{\dagger}e^{-i\varphi}\right]  +\\
i\hat{\rho}_{a}^{\dagger}\left[  \hat{e}^{\dagger}\hat{f}\hat{g}+\hat{e}%
\hat{f}\hat{g}^{\dagger}+\hat{e}\hat{f}^{\dagger}\hat{g}e^{-i\varphi}-\hat
{e}\hat{f}\hat{g}^{\dagger}e^{-i\varphi}\right]
\end{array}
\right\}  ,
\]%
\[
\hat{O}_{2}=\left(  \hat{\rho}_{b}^{\dagger}+i\hat{\rho}_{a}^{\dagger}\right)
\left(  2\hat{f}\hat{f}\hat{h}^{\dagger}+\hat{f}\hat{f}^{\dagger}\hat{h}%
+\hat{f}^{\dagger}\hat{f}\hat{h}\right)  ,
\]%
\[
\hat{O}_{3}=2\left\{
\begin{array}
[c]{c}%
\hat{\rho}_{a}^{\dagger}\left[  \hat{e}^{\dagger}\hat{f}\hat{h}-\hat{e}\hat
{f}\hat{h}^{\dagger}-\hat{e}\hat{f}^{\dagger}\hat{h}e^{-i\varphi}-\hat{e}%
\hat{f}\hat{h}^{\dagger}e^{-i\varphi}\right]  -\\
i\hat{\rho}_{b}^{\dagger}\left[  \hat{e}^{\dagger}\hat{f}\hat{h}-\hat{e}%
\hat{f}\hat{h}^{\dagger}+\hat{e}\hat{f}^{\dagger}\hat{h}e^{-i\varphi}+\hat
{e}\hat{f}\hat{h}^{\dagger}e^{-i\varphi}\right]
\end{array}
\right\}  ,
\]%
\[
\hat{O}_{4}=\left(  \hat{\rho}_{a}^{\dagger}-i\hat{\rho}_{b}^{\dagger}\right)
\left(  2\hat{f}\hat{f}\hat{g}^{\dagger}-\hat{f}\hat{f}^{\dagger}\hat{g}%
-\hat{f}^{\dagger}\hat{f}\hat{g}\right)  ,
\]%
\[
\hat{O}_{5}=\left(  \hat{\rho}_{b}^{\dagger}-i\hat{\rho}_{a}^{\dagger}\right)
\left(  2\hat{e}\hat{e}\hat{h}^{\dagger}-\hat{e}\hat{e}^{\dagger}\hat{h}%
-\hat{e}^{\dagger}\hat{e}\hat{h}\right)  e^{-i\varphi},
\]%
\[
\hat{O}_{6}=\left(  \hat{\rho}_{a}^{\dagger}+i\hat{\rho}_{b}^{\dagger}\right)
\left(  2\hat{e}\hat{e}\hat{g}^{\dagger}+\hat{e}\hat{e}^{\dagger}\hat{g}%
+\hat{e}^{\dagger}\hat{e}\hat{g}\right)  e^{-i\varphi};
\]
and
\begin{equation}
\left\langle \hat{D}_{\alpha,\varepsilon;\gamma,\varepsilon}\right\rangle
=\left[
\begin{array}
[c]{c}%
\left\langle \hat{M}_{1}\right\rangle +\left\langle \hat{M}_{1}^{\dagger
}\right\rangle +\left\langle \hat{M}_{2}\right\rangle +\left\langle \hat
{M}_{2}^{\dagger}\right\rangle \\
-2\{(\left\langle \hat{\rho}_{a}^{\dagger}\hat{g}\right\rangle +\left\langle
\hat{\rho}_{b}^{\dagger}\hat{h}\right\rangle )\left(  1-e^{-i\varphi}\right)
+(\left\langle \hat{\rho}_{a}\hat{g}^{\dagger}\right\rangle +\left\langle
\hat{\rho}_{b}\hat{h}^{\dagger}\right\rangle )\left(  1-e^{i\varphi}\right) \\
+i(\left\langle \hat{\rho}_{a}^{\dagger}\hat{h}\right\rangle -\left\langle
\hat{\rho}_{b}^{\dagger}\hat{g}\right\rangle )\left(  1+e^{-i\varphi}\right)
-i(\left\langle \hat{\rho}_{a}\hat{h}^{\dagger}\right\rangle -\left\langle
\hat{\rho}_{b}\hat{g}^{\dagger}\right\rangle )\left(  1+e^{i\varphi}\right)
\}\cdot\\
\{(\left\langle \hat{e}^{\dagger}\hat{g}\right\rangle +\left\langle \hat
{e}\hat{g}^{\dagger}\right\rangle -\left\langle \hat{f}^{\dagger}\hat
{h}\right\rangle -\left\langle \hat{f}\hat{h}^{\dagger}\right\rangle
)+i(\left\langle \hat{e}^{\dagger}\hat{h}\right\rangle -\left\langle \hat
{e}\hat{h}^{\dagger}\right\rangle -\left\langle \hat{f}^{\dagger}\hat
{g}\right\rangle +\left\langle \hat{f}\hat{g}^{\dagger}\right\rangle )\}
\end{array}
\right]  ,\nonumber
\end{equation}
where
\[
\hat{M}_{1}=\left\{
\begin{array}
[c]{c}%
\lbrack\hat{\rho}_{a}^{\dagger}(2\hat{e}^{\dagger}\hat{g}\hat{g}+\hat{e}%
\hat{g}\hat{g}^{\dagger}+\hat{e}\hat{g}^{\dagger}\hat{g}-2\hat{f}^{\dagger
}\hat{g}\hat{h}-2\hat{f}\hat{g}\hat{h}^{\dagger})+\\
\hat{\rho}_{b}^{\dagger}(2\hat{e}^{\dagger}\hat{g}\hat{h}+2\hat{e}\hat
{g}^{\dagger}\hat{h}-2\hat{f}^{\dagger}\hat{h}\hat{h}-\hat{f}\hat{h}\hat
{h}^{\dagger}-\hat{f}\hat{h}^{\dagger}\hat{h})]+\\
i[\hat{\rho}_{a}^{\dagger}(2\hat{e}^{\dagger}\hat{g}\hat{h}-2\hat{e}\hat
{h}^{\dagger}\hat{g}-2\hat{f}^{\dagger}\hat{g}\hat{g}+\hat{f}\hat{g}\hat
{g}^{\dagger}+\hat{f}\hat{g}^{\dagger}\hat{g})+\\
\hat{\rho}_{b}^{\dagger}(2\hat{e}^{\dagger}\hat{h}\hat{h}-\hat{e}\hat{h}%
\hat{h}^{\dagger}-\hat{e}\hat{h}^{\dagger}\hat{h}-2\hat{f}^{\dagger}\hat
{g}\hat{h}+2\hat{f}\hat{g}^{\dagger}\hat{h})]
\end{array}
\right\}  \left(  1-e^{-i\varphi}\right)  ,
\]%
\[
\hat{M}_{2}=\left\{
\begin{array}
[c]{c}%
\lbrack\hat{\rho}_{a}^{\dagger}(2\hat{f}^{\dagger}\hat{g}\hat{h}-2\hat{f}%
\hat{g}^{\dagger}\hat{h}-2\hat{e}^{\dagger}\hat{h}\hat{h}+\hat{e}\hat{h}%
\hat{h}^{\dagger}+\hat{e}\hat{h}^{\dagger}\hat{h})+\\
\hat{\rho}_{b}^{\dagger}(2\hat{e}^{\dagger}\hat{g}\hat{h}-2\hat{e}\hat{g}%
\hat{h}^{\dagger}-2\hat{f}^{\dagger}\hat{g}\hat{g}+\hat{f}\hat{g}\hat
{g}^{\dagger}+\hat{f}\hat{g}^{\dagger}\hat{g})]+\\
i[\hat{\rho}_{a}^{\dagger}(2\hat{e}^{\dagger}\hat{g}\hat{h}+2\hat{e}\hat
{g}^{\dagger}\hat{h}-2\hat{f}^{\dagger}\hat{h}\hat{h}-\hat{f}\hat{h}\hat
{h}^{\dagger}-\hat{f}\hat{h}^{\dagger}\hat{h})+\\
\hat{\rho}_{b}^{\dagger}(2\hat{f}^{\dagger}\hat{g}\hat{h}+2\hat{f}\hat{g}%
\hat{h}^{\dagger}-2\hat{e}^{\dagger}\hat{g}\hat{g}-\hat{e}\hat{g}\hat
{g}^{\dagger}-\hat{e}\hat{g}^{\dagger}\hat{g})]
\end{array}
\right\}  \left(  1+e^{-i\varphi}\right)  .
\]

\section{APPLICATIONS}

For the purpose of validation, it is demonstrated in this section that when
the expressions for the loss model given in the previous three sections are
used in conjunction with eq.(\ref{D2PHI}), they yield the required phase
sensitivity results that have been previously developed in PI for the lossless
MZI and the ground state MZI . As an additional illustrative application, this
model is also used to derive the phase sensitivity $\Delta^{2}\varphi_{\gamma
}$ for an MZI with its two internal arms (the "$\gamma$ region") in the
excited state $\left\vert \psi_{e}\right\rangle \left\vert \psi_{f}%
\right\rangle =\left\vert 1\right\rangle \left\vert 1\right\rangle $ and the
phase sensitivity $\Delta^{2}\varphi_{\varepsilon}$ for an MZI with its two
channels between the output ports and ideal detectors (the "$\varepsilon$
region") in the excited state $\left\vert \psi_{g}\right\rangle \left\vert
\psi_{h}\right\rangle =\left\vert 1\right\rangle \left\vert 1\right\rangle $.
These results are used to determine a condition which relates the expected
number of photons entering an MZI to its efficiency parameters, such that
$\Delta^{2}\varphi_{\gamma}>\Delta^{2}\varphi_{\varepsilon}$ when this
condition is satisfied.

\subsection{Phase sensitivity for a lossless MZI}

Although this case is trivial and only requires the evaluation of
eqs.(\ref{DCA}) and (\ref{VARIN}), it is included\ here for completeness. When
the MZI is lossless, then each of the regional efficiencies has unit value so
that - with the exception $\kappa_{\alpha}=\frac{1}{4}$ - all of the powers
and products of subscripted $\kappa$'s in eq.(\ref{D2PHI}) are zero valued,
$\widehat{\rho}_{a}=\widehat{a}_{in}$, and $\widehat{\rho}_{b}=\widehat
{b}_{in}$. In this case eqs.(\ref{CA}) and (\ref{VARIN}) reduce to the
quantities $\left\langle \widehat{C}_{\alpha}^{\left(  0\right)
}\right\rangle $ and $\Delta^{2}C_{\alpha}^{\left(  0\right)  }$ defined in
section IV of PI and eq.(\ref{D2PHI}) reduces to the expression for the phase
sensitivity $\Delta^{2}\varphi_{lossless}$ for a lossless MZI given by eq.(10) therein.

\subsection{Phase sensitivity for the ground state MZI}

The ground state MZI is defined when the regional efficiency parameters in the
MZI model have their values in the open real interval $\left(  0,1\right)  $
and the system state is $\left\vert \Psi_{gs}\right\rangle =\left\vert
\psi_{a_{in},b_{in}}\right\rangle \left\vert 0\right\rangle \left\vert
0\right\rangle \left\vert 0\right\rangle \left\vert 0\right\rangle \left\vert
0\right\rangle \left\vert 0\right\rangle $. In this case, eventhough each of
the subscripted $\kappa$'s in eq.(\ref{D2PHI}) is non-vanishing - with the
exception of the first, fourth, and fifth terms in the numerator and the first
term in the denominator - each term in eq.(\ref{D2PHI}) is zero when its value
is determined using $\left\vert \Psi\right\rangle =\left\vert \Psi
_{gs}\right\rangle $. Consequently, eq.(\ref{D2PHI}) reduces to the expression
for the ground state MZI phase sensitivity $\Delta^{2}\varphi_{gs}$ given by
eq.(11) in PI.

These terms are zero because they are sums of terms which vanish due to the
fact that each contains a factor $\left\langle \widehat{X}\right\rangle
_{vac}=0$, where $\widehat{X}$ is either a single environmental operator or a
juxtaposition of several like environmental annihilation or creation operators
and the subscript $"vac"$ refers to the fact that the mean value is evaluated
using only the associated environmental vacuum state. For example, $\Delta
^{2}C_{\gamma}=0$ because each term in eq.(\ref{VARFB}) vanishes, i.e. for the
first term%
\begin{align*}
\left\langle \widehat{e}^{\dag}\widehat{e}\widehat{f}\widehat{f}^{\dag
}\right\rangle  &  =\left\langle \Psi_{gs}\right\vert \widehat{e}^{\dag
}\widehat{e}\widehat{f}\widehat{f}^{\dag}\left\vert \Psi_{gs}\right\rangle \\
&  =\left\langle \psi_{a_{in},b_{in}}\right\vert \left.  \psi_{a_{in},b_{in}%
}\right\rangle \left\langle \psi_{c}\right\vert \left.  \psi_{c}\right\rangle
\left\langle \psi_{d}\right\vert \left.  \psi_{d}\right\rangle \left\langle
\psi_{e}\right\vert \widehat{e}^{\dag}\widehat{e}\left\vert \psi
_{e}\right\rangle \left\langle \psi_{f}\right\vert \widehat{f}\widehat
{f}^{\dag}\left\vert \psi_{f}\right\rangle \left\langle \psi_{g}\right\vert
\left.  \psi_{g}\right\rangle \left\langle \psi_{h}\right\vert \left.
\psi_{h}\right\rangle \\
&  =\left\langle \psi_{e}\right\vert \widehat{e}^{\dag}\widehat{e}\left\vert
\psi_{e}\right\rangle \left\langle \psi_{f}\right\vert \widehat{f}\widehat
{f}^{\dag}\left\vert \psi_{f}\right\rangle \\
&  =\left\langle 0\right\vert \widehat{e}^{\dag}\widehat{e}\left\vert
0\right\rangle \left\langle 0\right\vert \widehat{f}\widehat{f}^{\dag
}\left\vert 0\right\rangle \\
&  =0\cdot\left\langle 0\right\vert \widehat{f}\widehat{f}^{\dag}\left\vert
0\right\rangle \\
&  =0
\end{align*}
(the reader will also recognize $\widehat{e}^{\dag}\widehat{e}=\widehat{N}$ as
the number operator $\widehat{N}$ with the property $\left\langle 0\right\vert
\widehat{N}\left\vert 0\right\rangle =0$), and similarly for terms two through
seven. It is also easily verified that $\Delta^{2}C_{\alpha},\Delta
^{2}C_{\alpha,\gamma}\left(  =4\left\langle \Psi_{gs}\right\vert \left[
\widehat{F}_{3}+\widehat{F}_{6}\right]  \left\vert \Psi_{gs}\right\rangle
\right)  ,\Delta^{2}C_{\alpha,\varepsilon}\left(  =\left\langle \Psi
_{gs}\right\vert \left[  \widehat{R}_{3}+\widehat{R}_{6}\right]  \left\vert
\Psi_{gs}\right\rangle \right)  ,$ and $\frac{\partial\left\langle \widehat
{C}_{\alpha}\right\rangle }{\partial\varphi}$ are as specified in section V of PI.

\subsection{Phase sensitivities for simple excited state MZIs : A sensitivity
trade-off condition}

Consider now the phase sensitivities $\Delta^{2}\varphi_{\gamma}$ and
$\Delta^{2}\varphi_{\varepsilon}$ for MZIs in the system states $\left\vert
\Psi_{\gamma}\right\rangle =\left\vert \psi_{a_{in},b_{in}}\right\rangle
\left\vert 0\right\rangle \left\vert 0\right\rangle \left\vert 1\right\rangle
\left\vert 1\right\rangle \left\vert 0\right\rangle \left\vert 0\right\rangle
$ and $\left\vert \Psi_{\varepsilon}\right\rangle =\left\vert \psi
_{a_{in},b_{in}}\right\rangle \left\vert 0\right\rangle \left\vert
0\right\rangle \left\vert 0\right\rangle \left\vert 0\right\rangle \left\vert
1\right\rangle \left\vert 1\right\rangle $, respectively. When $\left\vert
\Psi_{\gamma}\right\rangle $ is used as the model's system state it is found
that all but the first, second, fourth, fifth, and sixth terms in the
numerator and the first term in the denominator of eq.(\ref{D2PHI}) vanish.
Using eq.(\ref{VARIF}) it is also found that%
\[
\Delta^{2}C_{\alpha,\gamma}=\Delta^{2}C_{\alpha,\gamma}^{gs}+\Delta
^{2}C_{\alpha,\gamma}^{\gamma},
\]
where%
\[
\Delta^{2}C_{\alpha,\gamma}^{gs}\equiv4\left\langle \Psi_{\gamma}\right\vert
\left[  \widehat{F}_{3}+\widehat{F}_{6}\right]  \left\vert \Psi_{\gamma
}\right\rangle =4\left\langle \Psi_{gs}\right\vert \left[  \widehat{F}%
_{3}+\widehat{F}_{6}\right]  \left\vert \Psi_{gs}\right\rangle
\]
and%
\[
\Delta^{2}C_{\alpha,\gamma}^{\gamma}\equiv4\left\langle \Psi_{\gamma
}\right\vert \left[  \widehat{F}_{2}+\widehat{F}_{3}+\widehat{F}_{5}%
+\widehat{F}_{6}\right]  \left\vert \Psi_{\gamma}\right\rangle =4\left[
\alpha\left(  \left\langle \widehat{a}_{in}^{\dag}\widehat{a}_{in}%
\right\rangle +\left\langle \widehat{b}_{in}^{\dag}\widehat{b}_{in}%
\right\rangle \right)  +1\right]  \text{.}%
\]
Since\ each of the expressions for $\Delta^{2}C_{\alpha},\Delta^{2}%
C_{\alpha,\varepsilon},$ and $\frac{\partial\left\langle \widehat{C}_{\alpha
}\right\rangle }{\partial\varphi}$ remains invariant when the system states
$\left\vert \Psi_{gs}\right\rangle $ and $\left\vert \Psi_{\gamma
}\right\rangle $ are used for their valuations, then%
\[
\Delta^{2}\varphi_{\gamma}=\Delta^{2}\varphi_{gs}+\Delta^{2}\varphi_{\gamma
}^{e}\text{,}%
\]
where%
\[
\Delta^{2}\varphi_{gs}\equiv\frac{\kappa_{\alpha}^{2}\Delta^{2}C_{\alpha
}+\kappa_{\alpha,\gamma}^{2}\Delta^{2}C_{\alpha,\gamma}^{gs}+\kappa
_{\alpha,\varepsilon}^{2}\Delta^{2}C_{\alpha,\varepsilon}}{\left\vert
\kappa_{\alpha}\frac{\partial\left\langle \widehat{C}_{\alpha}\right\rangle
}{\partial\varphi}\right\vert ^{2}}%
\]
is the phase sensitivity for the ground state MZI and%
\[
\Delta^{2}\varphi_{\gamma}^{e}\equiv\frac{\kappa_{\gamma}^{2}\Delta
^{2}C_{\gamma}+\kappa_{\alpha,\gamma}^{2}\Delta^{2}C_{\alpha,\gamma}^{\gamma
}+\kappa_{\gamma,\varepsilon}^{2}\Delta^{2}C_{\gamma,\varepsilon}}{\left\vert
\kappa_{\alpha}\frac{\partial\left\langle \widehat{C}_{\alpha}\right\rangle
}{\partial\varphi}\right\vert ^{2}}.
\]
Evaluation of the right hand side of the last equation yields (this corrects
the expression for $\Delta^{2}\varphi_{es}$ given in section VI of PI)%
\begin{equation}
\Delta^{2}\varphi_{\gamma}=\Delta^{2}\varphi_{gs}+\left[  \frac{2\left(
1-\gamma\right)  }{\alpha^{2}\gamma^{2}\varepsilon}\right]  \left\{
\frac{\varepsilon\left[  \alpha\gamma\left(  \left\langle \widehat{a}%
_{in}^{\dag}\widehat{a}_{in}\right\rangle +\left\langle \widehat{b}_{in}%
^{\dag}\widehat{b}_{in}\right\rangle \right)  +\left(  1-\gamma\right)
\right]  +1}{\left[  \left(  \left\langle \widehat{b}_{in}^{\dag}\widehat
{b}_{in}\right\rangle -\left\langle \widehat{a}_{in}^{\dag}\widehat{a}%
_{in}\right\rangle \right)  \sin\varphi+\left(  \left\langle \widehat{a}%
_{in}^{\dag}\widehat{b}_{in}\right\rangle +\left\langle \widehat{b}_{in}%
^{\dag}\widehat{a}_{in}\right\rangle \right)  \cos\varphi\right]  ^{2}%
}\right\}  \text{.} \label{GSEN}%
\end{equation}
Thus, if the "$\gamma$ region" is in the excited state $\left\vert
1\right\rangle \left\vert 1\right\rangle $ and $\gamma\in\left(  0,1\right)
$, then $\Delta^{2}\varphi_{\gamma}>\Delta^{2}\varphi_{gs}$. Observe that when
there are no losses in the "$\gamma$ region" , i.e. when $\gamma=1$, then - as
required - $\Delta^{2}\varphi_{\gamma}^{e}=0$ so that $\Delta^{2}%
\varphi_{\gamma}=\Delta^{2}\varphi_{gs}$.

When $\left\vert \Psi_{\varepsilon}\right\rangle $ is used as the model's
system state, then all terms in eq.(\ref{D2PHI}) vanish except the first term
in the denominator and the first, fourth, fifth, and sixth terms in the
numerator. Also, it is determined from eq.(\ref{VARID}) that
\[
\Delta^{2}C_{\alpha,\varepsilon}=\Delta^{2}C_{\alpha,\varepsilon}^{gs}%
+\Delta^{2}C_{\alpha,\varepsilon}^{\varepsilon},
\]
where%
\[
\Delta^{2}C_{\alpha,\varepsilon}^{gs}\equiv\left\langle \Psi_{\varepsilon
}\right\vert \left[  \widehat{R}_{3}+\widehat{R}_{6}\right]  \left\vert
\Psi_{\varepsilon}\right\rangle =\left\langle \Psi_{gs}\right\vert \left[
\widehat{R}_{3}+\widehat{R}_{6}\right]  \left\vert \Psi_{gs}\right\rangle
\]
and%
\[
\Delta^{2}C_{\alpha,\varepsilon}^{\varepsilon}\equiv\left\langle
\Psi_{\varepsilon}\right\vert \left[  \widehat{R}_{2}+\widehat{R}_{3}%
+\widehat{R}_{5}+\widehat{R}_{6}\right]  \left\vert \Psi_{\varepsilon
}\right\rangle =8\left[  \alpha\left(  \left\langle \widehat{a}_{in}^{\dag
}\widehat{a}_{in}\right\rangle +\left\langle \widehat{b}_{in}^{\dag}%
\widehat{b}_{in}\right\rangle \right)  +1\right]  ,
\]
and it is easily verified that the quantities $\Delta^{2}C_{\alpha},\Delta
^{2}C_{\alpha,\gamma},$ and $\frac{\partial\left\langle \widehat{C}_{\alpha
}\right\rangle }{\partial\varphi}$ yield identical expressions when the system
states $\left\vert \Psi_{gs}\right\rangle $ and $\left\vert \Psi_{\varepsilon
}\right\rangle $ are used to evaluate them. Thus,%
\[
\Delta^{2}\varphi_{\varepsilon}=\Delta^{2}\varphi_{gs}+\Delta^{2}%
\varphi_{\varepsilon}^{e},
\]
where%
\[
\Delta^{2}\varphi_{gs}\equiv\frac{\kappa_{\alpha}^{2}\Delta^{2}C_{\alpha
}+\kappa_{\alpha,\gamma}\Delta^{2}C_{\alpha,\gamma}+\kappa_{\alpha
,\varepsilon}^{2}\Delta^{2}C_{\alpha,\varepsilon}^{gs}}{\left\vert
\kappa_{\alpha}\frac{\partial\left\langle \widehat{C}_{\alpha}\right\rangle
}{\partial\varphi}\right\vert ^{2}}%
\]
is the phase sensitivity for the ground state MZI and%
\[
\Delta^{2}\varphi_{\varepsilon}^{e}\equiv\frac{\kappa_{\alpha,\varepsilon}%
^{2}\Delta^{2}C_{\alpha,\varepsilon}^{\varepsilon}+\kappa_{\gamma,\varepsilon
}^{2}\Delta^{2}C_{\gamma,\varepsilon}}{\left\vert \kappa_{\alpha}%
\frac{\partial\left\langle \widehat{C}_{\alpha}\right\rangle }{\partial
\varphi}\right\vert ^{2}},
\]
so that%
\begin{equation}
\Delta^{2}\varphi_{\varepsilon}=\Delta^{2}\varphi_{gs}+\left[  \frac{2\left(
1-\varepsilon\right)  }{\alpha^{2}\gamma^{2}\varepsilon}\right]  \left\{
\frac{\alpha\gamma\left(  \left\langle \widehat{a}_{in}^{\dag}\widehat{a}%
_{in}\right\rangle +\left\langle \widehat{b}_{in}^{\dag}\widehat{b}%
_{in}\right\rangle \right)  +1}{\left[  \left(  \left\langle \widehat{b}%
_{in}^{\dag}\widehat{b}_{in}\right\rangle -\left\langle \widehat{a}_{in}%
^{\dag}\widehat{a}_{in}\right\rangle \right)  \sin\varphi+\left(  \left\langle
\widehat{a}_{in}^{\dag}\widehat{b}_{in}\right\rangle +\left\langle \widehat
{b}_{in}^{\dag}\widehat{a}_{in}\right\rangle \right)  \cos\varphi\right]
^{2}}\right\}  . \label{ESEN}%
\end{equation}
It is clear from this expression that when the "$\varepsilon$ region" is in
the excited state $\left\vert 1\right\rangle \left\vert 1\right\rangle $ and
$\varepsilon\in\left(  0,1\right)  $, then $\Delta^{2}\varphi_{\varepsilon
}>\Delta^{2}\varphi_{gs}$ and when there are no losses in the "$\varepsilon$
region" so that $\varepsilon=1$, then - as required - $\Delta^{2}%
\varphi_{\varepsilon}^{e}=0$ and $\Delta^{2}\varphi_{\varepsilon}=\Delta
^{2}\varphi_{gs}$.

For the sake of further illustrating the utility of the model, note that
eqs.(\ref{GSEN}) and (\ref{ESEN}) yield the difference
\[
\Delta^{2}\varphi_{\gamma}-\Delta^{2}\varphi_{\varepsilon}=\frac{2\alpha
\gamma\left[  \varepsilon\left(  1-\gamma\right)  -\left(  1-\varepsilon
\right)  \right]  \left(  \left\langle \widehat{a}_{in}^{\dag}\widehat{a}%
_{in}\right\rangle +\left\langle \widehat{b}_{in}^{\dag}\widehat{b}%
_{in}\right\rangle \right)  +2\left[  \varepsilon\left(  1-\gamma\right)
^{2}+\varepsilon-\gamma\right]  }{\alpha^{2}\gamma^{2}\varepsilon\left[
\left(  \left\langle \widehat{b}_{in}^{\dag}\widehat{b}_{in}\right\rangle
-\left\langle \widehat{a}_{in}^{\dag}\widehat{a}_{in}\right\rangle \right)
\sin\varphi+\left(  \left\langle \widehat{a}_{in}^{\dag}\widehat{b}%
_{in}\right\rangle +\left\langle \widehat{b}_{in}^{\dag}\widehat{a}%
_{in}\right\rangle \right)  \cos\varphi\right]  ^{2}}.
\]
Thus, for $\gamma\neq1$ or $\varepsilon\neq1$ it can be concluded that when
the condition%
\begin{equation}
\alpha\left(  \left\langle \widehat{a}_{in}^{\dag}\widehat{a}_{in}%
\right\rangle +\left\langle \widehat{b}_{in}^{\dag}\widehat{b}_{in}%
\right\rangle \right)  >\frac{\left[  \gamma-\varepsilon-\varepsilon\left(
1-\gamma\right)  ^{2}\right]  }{\gamma\left[  \varepsilon\left(
1-\gamma\right)  -\left(  1-\varepsilon\right)  \right]  } \label{INEQ}%
\end{equation}
is satisfied, then $\Delta^{2}\varphi_{\gamma}>\Delta^{2}\varphi_{\varepsilon
}$, i.e. \emph{phase sensitivity is more degraded when an MZI is in state}
$\left\vert \Psi_{\gamma}\right\rangle $ \emph{than when it is in state}
$\left\vert \Psi_{\varepsilon}\right\rangle $. This is an intuitively pleasing
conclusion, since - unlike the case for the excited "$\varepsilon$ region" -
the degradation in sensitivity due to the excited "$\gamma$ region" would be
expected to experience additional degradation induced by the (unexcited)
"$\varepsilon$ region". As a special case of this result, observe that if
$\gamma=\varepsilon\neq1$, then - since the right hand side of (\ref{INEQ})
has unit value - the relationship $\Delta^{2}\varphi_{\gamma}>\Delta
^{2}\varphi_{\varepsilon}$ also prevails when%
\[
\left\langle \widehat{a}_{in}^{\dag}\widehat{a}_{in}\right\rangle
+\left\langle \widehat{b}_{in}^{\dag}\widehat{b}_{in}\right\rangle >\frac
{1}{\alpha}.
\]

\section{CLOSING\ REMARKS}

The above provides a generalized analytical tripartite loss model for MZI
phase sensitivity which is valid for both arbitrary photon input states and
arbitrary system environmental states. This model subsumes the phase
sensitivity models for the lossless MZI and the ground state MZI\ and is
useful for developing specialized models for estimating the phase sensitivity,
as well as for performing associated design trade-off analyses, for MZIs and
MZI-like instruments which operate in environmental regimes which are not
contained within the envelope of validity for the ground state model.

\end{document}